\documentclass[pra,aps,twocolumn,superscriptaddress,longbibliography,nofootinbib,floatfix]{revtex4-2}
\usepackage[portrait, margin=1in]{geometry}
\usepackage{amsmath,graphicx,epsfig}
\usepackage{epstopdf}
\usepackage{euscript}
\usepackage{amsfonts}
\usepackage{amssymb}
\usepackage[colorlinks]{hyperref}
\usepackage{tabularx}
\usepackage{bm}
\newcolumntype{Y}{>{\centering\arraybackslash}X}
\usepackage{xcolor}
\newcommand{\APD}[1]{{\color{black}{#1}}}

\def\bra#1{{\langle#1|}}

\def\cg(#1,#2)(#3,#4)(#5,#6){\bra{#1,#2,#3,#4}#5,#6\rangle}

\def\threej(#1,#2)(#3,#4)(#5,#6){\begin{pmatrix}#1&#3&#5\\#2&#4&#6\end{pmatrix}}
\def\sixj(#1,#2,#3)(#4,#5,#6){\begin{Bmatrix}#1&#2&#3\\#4&#5&#6\end{Bmatrix}}
\def\ninej(#1,#2,#3)(#4,#5,#6)(#7,#8,#9){\begin{Bmatrix}#1&#2&#3\\#4&#5&#6\\#7&#8&#9\end{Bmatrix}}

\def\mr{\mathrm}
\def\mb{\mathbf}
\def\bs{\boldsymbol}

\def\acr{{\ensuremath{\alpha^\star}}}

\graphicspath{{Figures/}}
\begin{document}
\title{Anthropic constraint on transient variations of fundamental constants}
\author{V. D. Dergachev}
 \affiliation{Department of Chemistry, University of Nevada, Reno, Nevada 89557, USA}

\author{H. B. Tran Tan}
\affiliation{Department of Physics, University of Nevada, Reno, Nevada 89557, USA}

\author{S. A. Varganov}
\affiliation{Department of Chemistry, University of Nevada, Reno, Nevada 89557, USA}

\author{A. Derevianko}
\affiliation{Department of Physics, University of Nevada, Reno, Nevada 89557, USA}
\begin{abstract}
    The anthropic principle implies that life can emerge and be sustained only in a narrow range of values of fundamental constants. Here we show that anthropic arguments can set powerful constraints on {\em transient} variations of the fine-structure constant $\alpha$ over the past 4 billion years since the appearance of lifeforms on Earth. 
    We argue that the passage through Earth of a macroscopic dark matter clump with a value of $\alpha$ inside differing substantially from its nominal value would make Earth uninhabitable. We demonstrate that in the regime of extreme variation of $\alpha$,  the periodic table of elements is truncated, water fails to serve as a universal solvent, and protons become unstable.
    Thereby, the anthropic principle constrains the likelihood of such encounters on a 4-billion-year timescale. This enables us to improve existing astrophysical bounds on certain dark  matter model couplings by several orders of magnitude.  
\end{abstract}
\maketitle

The anthropic principle implies that life as we know it can emerge and be sustained only in a certain range of fundamental constants (FCs)~\cite{Brandon_Carter,carr1979anthropic,Hogan1999}. While the values of FCs are fixed in the Standard Model of elementary particles, modern theories promote them to dynamic quantities. Constants are no longer constant. 
\APD{The bulk of the literature on FCs variation implicitly assumes that FCs differ from their nominal values over vast cosmological volumes. In this regime, stringent constraints on variation of FCs do exist~\cite{WebKinMur11-dipole,NIST_Science_clock,ptb_atomic_clock,Oklo_Lamoreaux}. However, these constraints can be evaded when the FCs differ from their nominal values only over volumes (clumps) much smaller than the galactic scales. Moving clumps translate into transient variations of FCs in the observer’s frame. Transient variations are motivated by clumpy dark matter (DM) models such as topological defects, Q-balls or dark stars~\cite{VILENKIN1985263,Kusenko2001,DerPos14,Braaten2019}. In these models, the FCs inside and outside the DM clumps can differ substantially. 

As an illustration of how the existing constraints can be evaded, consider the celebrated triple-$\alpha$-particle process synthesizing  $^{12}\mathrm{C}$ nuclei. The production rate is strongly enhanced by a fortuitously placed Hoyle resonance~\cite{hoyle1954nuclear,Dunbar1953C12}. It is commonly argued (see, e.g., Ref.~\cite{barrow1996jd,Lahde2020}) that if the fine-structure constant $\alpha$ were off by a few percent, the shift in the resonance energy would inhibit stellar production of $^{12}\mathrm{C}$ required for the emergence of carbon-based life. This anthropic argument presupposes that every star is affected by the variation of $\alpha$. If, however, $\alpha$ differs from its nominal value $\alpha_0$ only for a brief period of time inside some star (due to a clump sweep),  $^{12}\mathrm{C}$ would be still produced either in other stars or outside the encounter window. Similarly, tight astrophysical bounds on FC drifts are derived from spectroscopic observations of gas clouds seen in absorption spectra against background quasars~\cite{WebKinMur11-dipole}. However, if the DM clump does not overlap with the gas cloud along the line of sight and at the right moment, these  constraints are, once again, evaded by the  transients.}

So far the direct observational constraints on transient variations of FCs come from atomic clocks~\cite{NIST_Science_clock,ptb_atomic_clock}. Encounters of Earth with DM clumps can be exceedingly rare, while the direct searches only extend over a short 20-year recent history~\cite{Roberts2017-GPS-DM}. Here, we show that anthropic arguments can set powerful constraints on FC transients over the past 4 billion years since the appearance of lifeforms on Earth.

Life primarily depends on the elements in the first two rows of the Mendeleev periodic table. If a DM clump envelopes  Earth and substantially affects the properties of these elements and the molecules containing them, it can dramatically change the conditions for life. At the quantum level, biology mainly depends on a set of three FCs: the electron mass $m_e$, the elementary charge $e$, and the Planck constant $\hbar$. Variations in these FCs lead to a trivial isotropic scaling of the molecular geometries preserving bond angles, thus having no effects on biology, see Appendix A. However, this changes when relativity brings in the speed of light $c$ or, equivalently, $\alpha=e^2/\left(\hbar c\right)$. As we show, relativistic effects overturn this isotropic scaling, dramatically modifying molecular geometries. Substantial changes occur at extreme values of  $\alpha/\alpha_0=c_0/c\sim 10$, where $\alpha_0\approx1/137$ and $c_0$ are the nominal values.

\APD{Considering that the variations in FCs for exceedingly rare transients are not constrained, here we explore a  novel regime of {\em large} variations. 
While it is intuitively clear that a large variation in FCs can be disastrous for life, the detailed reasons  are scantily understood and we address them here.}
We find an abundance of remarkable effects on the structure of atoms and molecules and, by extension, on the fundamental conditions for the emergence and sustainability of life. For $\alpha\sim10\alpha_0$, the periodic table shrinks to elements from hydrogen to sulphur, the aufbau principle of atomic structure is qualitatively changed, noble gasses are no longer inert,  water fails to serve as a universal solvent, 
\APD{and protons become unstable.} If a DM clump with $\alpha\sim 10\alpha_0$ were to envelope Earth, it would have been catastrophic for life. 

\begin{figure}
    \centering
    \includegraphics[scale=0.225]{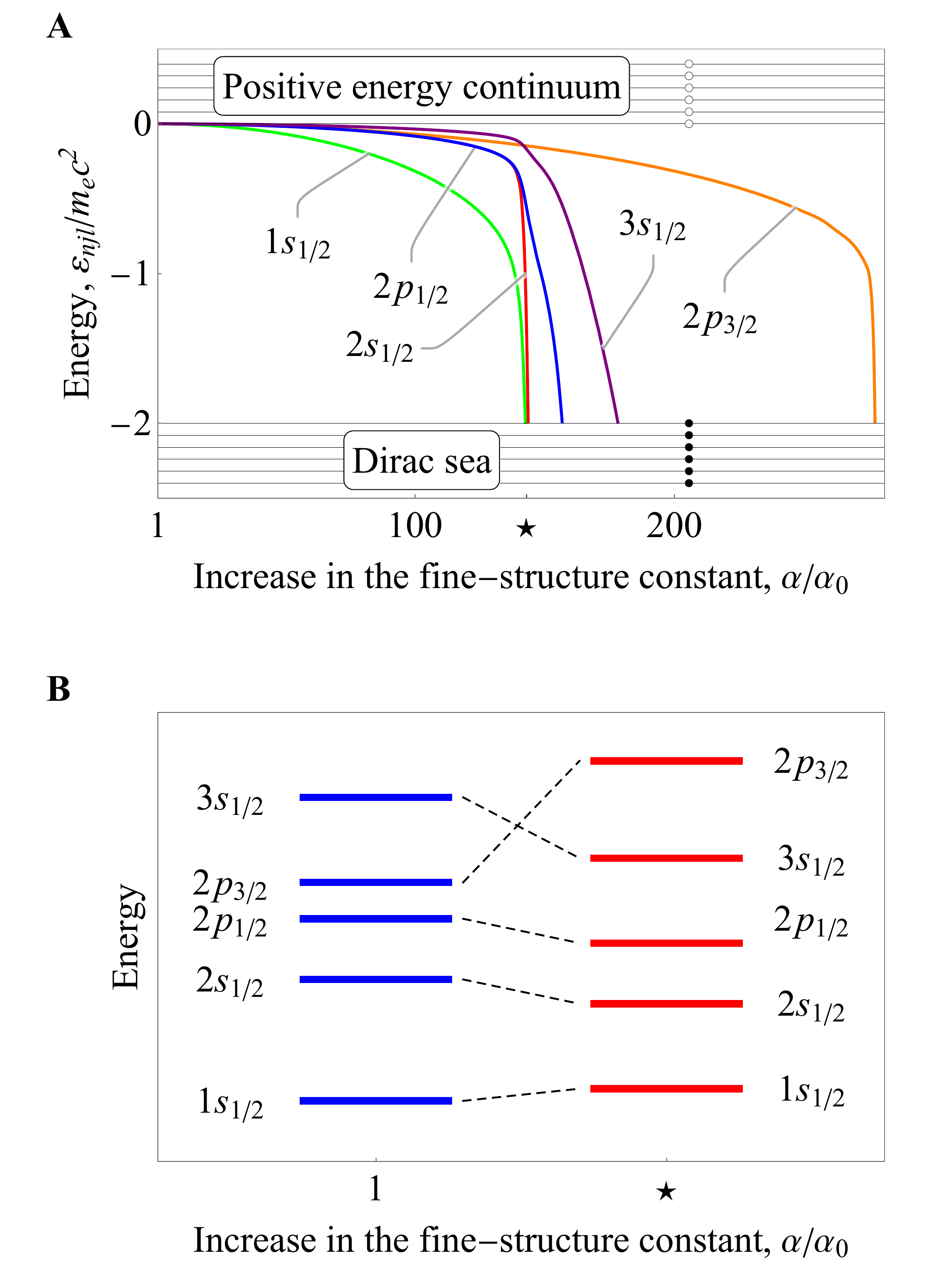}
    \caption{Dependence of atomic energy levels on ${\alpha}$. \textbf{(A)} Energy levels of atomic hydrogen as functions of $\alpha$. The $1s_{1/2}$ level dives into the Dirac sea at $\alpha^\star\approx143\alpha_0$. \textbf{(B)} Evolution of the aufbau principle’s sequence of shell occupation in many-electron atoms with varying $\alpha$.}
    \label{fig:Dirac}
\end{figure}

\textit{Effects of large $\alpha$ variations ---} Qualitatively, an increase in $\alpha$ can be mapped into a reduction in the apparent value of speed of light $c$, amplifying relativistic effects. To properly describe this ultra-relativistic regime, we solve the four-component Dirac equation~\cite{Reiher2015} using the DIRAC package~\cite{DIRAC21} and other non-perturbative-in-$\alpha$ methods.
We start with the hydrogen ion with a point-like nucleus. The ground state energy is given by
\begin{equation}
    \varepsilon_{{1s}_{1/2}}=m_ec^2\left(\sqrt{1-\alpha^2}-1\right)\,.
\end{equation}
As $\alpha$ increases, the energy is lowered towards the lower continuum (Dirac sea, Fig.~\ref{fig:Dirac}) until $\alpha$ reaches 1 where $\varepsilon_{{1s}_{1/2}}=-m_ec^2$. For $\alpha>1$, the argument of the square root becomes negative and $\varepsilon_{{1s}_{1/2}}$ acquires an imaginary part: the ground state becomes unstable. A more realistic analysis with a finite-sized nucleus shows that the ${1s}_{1/2}$ energy ``dives'' into the Dirac sea at $-{2m}_ec^2$ at the critical value of $\alpha^\star\approx1.04=143\alpha_0$, see Appendix B. For $\alpha>\alpha^\star$, the discrete ${1s}_{1/2}$ level becomes embedded into the Dirac sea continuum. Then an electron-positron pair is emitted spontaneously and the vacuum becomes electrically charged, as discussed in the context of determining the critical nuclear charge with $\alpha$ fixed to its nominal value~\cite{ZeldovichPopov1972,GreRei02QED_book}.
For hydrogen-like ions, the critical value $\alpha^\star$ increases with its nuclear charge $Z$ as (see Appendix B)
\begin{equation}
    \alpha_0/\alpha^\star\approx Z/168\,.\label{eq:alphaZ}
\end{equation}
This remains a good approximation for multi-electron systems as the ${1s}_{1/2}$ electrons tend to experience the unscreened nuclear charge, with minor correlation corrections to binding energies. In a molecule, $\alpha^\star$ is determined by the charge of the heaviest nucleus: for water  $\alpha_\mathrm{H_2O}^\star\approx \alpha_\mathrm{O}^\star\approx18.4\alpha_0$ is fixed by the oxygen atom. Eq.~\eqref{eq:alphaZ} shows that for a given $\alpha^\star$, only elements with $Z\lesssim168\alpha_0/\alpha$ are stable: if $\alpha$ increases ten-fold, the entire periodic table shrinks to elements from hydrogen to sulphur. 

In 1939, George Gamow published the book ``Mr. Tompkins in Wonderland''~\cite{Gamow1939-MrTompkins} which tells a story about a world where FCs have values radically different from those they have in the real world. In Mr. Tompkins’ alternative reality, where $c$ is reduced to that of a speeding bicycle, $c/c_0=\alpha_0/\alpha\approx4\times{10}^{-8}$, even the hydrogen atom fails to exist.

One of the most powerful rules in atomic structure is the {\em aufbau} (building-up) principle which determines the sequence in which atomic shells are filled with electrons. For $\alpha\approx\alpha_0$, the textbook sequence is $1s_{1/2}2s_{1/2}2p_{1/2}2p_{3/2}3s_{1/2}$. In the ultra-relativistic regime ($\alpha\rightarrow\alpha^\star$), our calculations show that the energy of the $3s_{1/2}$ orbital drops below that of $2p_{3/2}$, changing the filling order of these shells. Qualitatively, the $2p_{3/2}$ orbital dives into the Dirac sea at $\alpha$ substantially larger than that for $2p_{1/2}$, see Fig.~\ref{fig:Dirac}. This leads to a giant $\sim m_ec^2$ fine-structure splitting near the critical value $\alpha^\star$. For the same reason, there is a large difference in the relativistic contraction of the $2p_{3/2}$ and $2p_{1/2}$ shells near $\alpha^\star$: the $2p_{1/2}$ (and the $3s_{1/2}$) shells become submerged inside the $2p_{3/2}$ shell. This drives a more effective screening of the nuclear charge by the inner shells, causing an increase in the $2p_{3/2}$ orbital energies with increasing $\alpha$ in many-electron atoms, a trend opposite to that in the H-like ions.

Our numerical solutions to the Dirac equation for second-period atoms with electrons in the $2p_{3/2}$ shell (N, O, F, and Ne) are consistent with the described modified {\em aufbau} principle of Fig.~\ref{fig:Dirac}. All the enumerated atoms demonstrate non-trivial changes in their ground states near their respective $\alpha^\star$.  In particular, the ground state of neon ($Z=10$), for $\alpha\lesssim14\alpha_0$, becomes $1s_{1/2}^22s_{1/2}^22p_{1/2}^23s_{1/2}^22p_{3/2}^2$ with the total angular momentum $J=2$ instead of the nominal closed-shell $J=0$ state. Neon is no longer inert. The valence-shell configuration of ultra-relativistic neon, $3s_{1/2}^22p_{3/2}^2$, closely resembles that of carbon at nominal $\alpha$, $2s^22p^2$. Such neon is expected to have as rich a chemistry as carbon. It is intriguing to imagine polymers, nanostructures, and an entirely new biology where neon plays the traditional role of carbon.

How would the extreme variation of $\alpha$ affect the structure and properties of molecules? Here we address this question by focusing on water. All known lifeforms use water as a universal solvent for various chemicals and as an essential component of many metabolic processes~\cite{Benner2004}. 
At the nominal value of $\alpha$, the water molecule is bent leading to a non-zero dipole moment essential to its chemical and biological properties. At $\alpha\approx18\alpha_0$,  $\mathrm{H}_2 \mathrm{O}$ becomes linear and as such has a vanishing dipole moment; water fails to serve as the universal solvent.
Our  calculations show that when $\alpha$ increases, the $\mathrm{H}_2 \mathrm{O}$  geometry changes drastically (Fig.~\ref{fig:MO} A – C). At $\alpha\approx14\alpha_0$, the bond angle contracts from ${104.5}^\mathrm{o}$ to ${90}^\mathrm{o}$. At larger $\alpha$ but still below $\alpha_{\rm O}^\star \approx18.4\alpha_0$, the water molecule becomes linear, consistent with Ref.~\cite{Dubillard2006}. This is due to the relativistic stabilization of the $2s_{1/2}$ and $3s_{1/2}$ orbitals with respect to the $2p_j$ orbitals of the oxygen atom, and the increased $2p_j$ fine-structure splitting. The formation of molecular orbitals (MOs) from atomic orbitals requires energy resonances and overlaps between the constituent atomic orbitals. Amplified relativity affects (i) the resonances via the changes in energies and (ii) the overlaps via the varying degrees of contraction of atomic orbitals, see Appendix D. The changes in the $\mathrm{H}_2 \mathrm{O}$  geometry  can also be understood in terms of the textbook valence-shell electron-pair repulsion (VSEPR) model, see Appendix D. 

The striking changes in the $\mathrm{H}_2 \mathrm{O}$ geometry at increased $\alpha$ would lead to alternative chemistry and biology. For example, in contrast to the bent water molecules that form three-dimensional networks of hydrogen bonds, the ultra-relativistic linear $\mathrm{H}_2 \mathrm{O}$  could only form two-dimensional networks. This is anticipated to substantially decrease the freezing and boiling points of water\cite{Stillnger1980,Chang2005}.  

Finally, we require that protons remain stable. Nominally, proton is lighter than neutron, precluding proton decay. The $p-n$ mass difference results from a delicate cancellation between electromagnetic ($\propto\alpha$) and $u-d$ quark mass splitting effects~\cite{Borsanyi2015}. Increasing $\alpha$ to $\sim 10\alpha_0$ spoils this balance leading \APD{ both to the $\beta^+$ decay $p\rightarrow n+e^+  + \nu_e$ and atomic electron capture $p+e^- \rightarrow n + \nu_e$}, both energetically forbidden at the nominal $\alpha$ but allowed at large $\alpha$, see Appendix F for  details. 

\begin{widetext}
 \begin{center}
     \begin{figure}
         \includegraphics[scale=0.75]{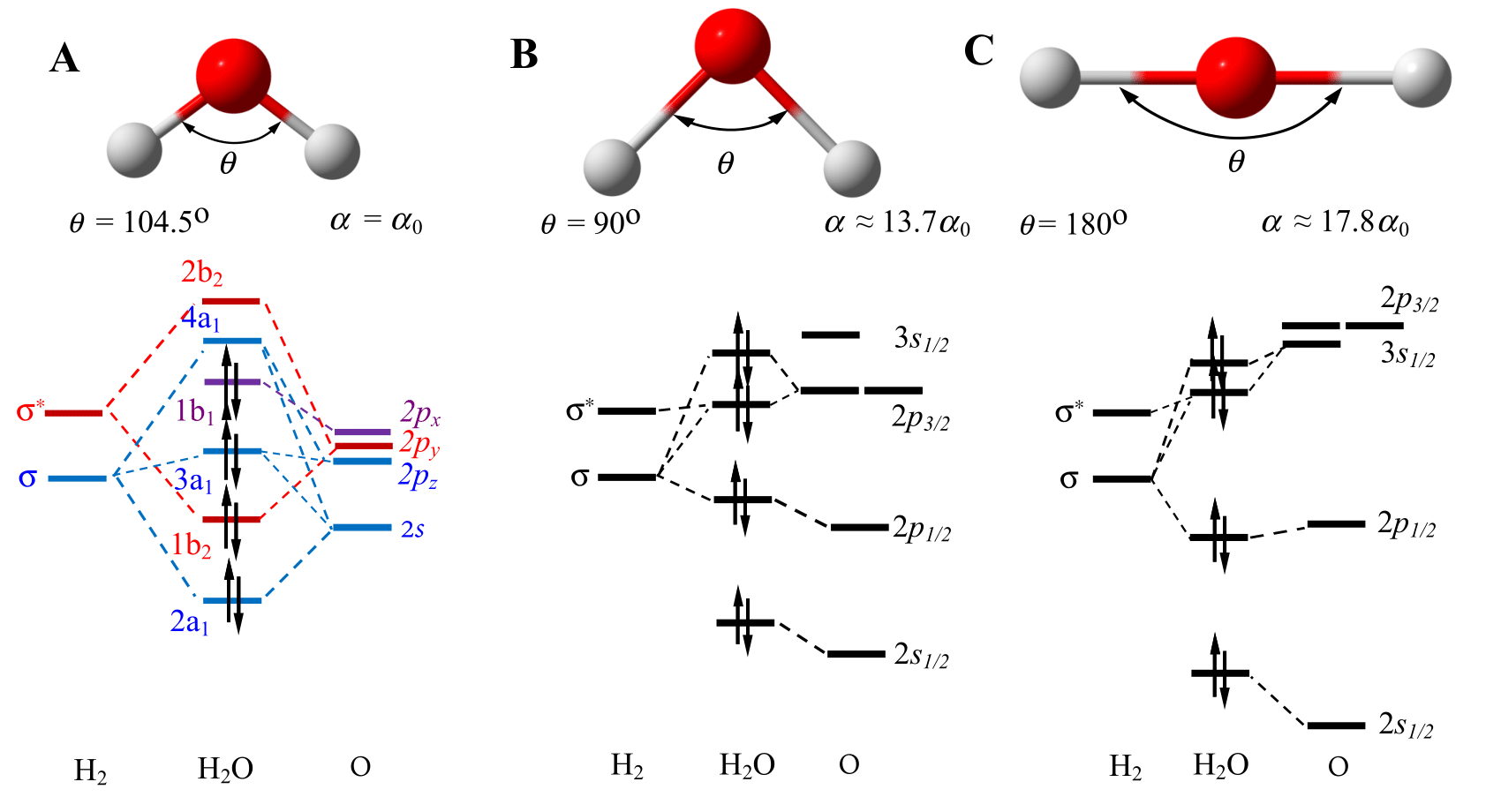}
         \caption{Molecular geometries and orbital diagrams of water at different $\alpha$.  At the nominal $\alpha$, the orbitals are labeled and colored according to their irreducible representations in the $C_{2v}$ point group. At $\alpha>\alpha_0$, the spin-orbital notation is used for atomic orbitals, and only the occupied molecular orbitals are shown. See Appendix D for details. }
         \label{fig:MO}
     \end{figure}
 \end{center}
\end{widetext}

\APD{ 
\textit{Anthropic bound on clumpy dark matter ---} 
While there are convincing astrophysical evidences for DM~\cite{RevModPhys.90.045002}, its nature remains a mystery.
Direct searches for the prominent DM candidates, the Weakly Interacting Massive Particles (WIMPs)~\cite{angloher2014,Aalseth2013,Armengaud2016,Bernabei2018,Agnese2018,Aprile2018,Akerib2019,Wang2020} and the axion~\cite{PRESKILL1983,ABBOTT1983,DINE1983,1475-7516-2013-07-013,Collaboration2017,GATTONE199959,MORALES2002325, PhysRevLett.112.241302,PhysRevD.84.121302,PhysRevLett.118.061302,McAllister:2017lkb,EHRET2010149,PhysRevD.78.092003,PhysRevLett.100.080402,DellaValle2016,QandA2007}, are inconclusive. 
This motivated a plethora of alternative DM candidates. If DM is formed from some underlying field $\phi$, 
these candidates can be separated into two classes: self-interacting and free-particle-like. Self-interaction potentials $U(\phi)$ often lead to formation of macroscopic DM clumps which can solve some outstanding difficulties with free-particle DM models, such as the galactic core-cusp problem~\cite{Newman_2013b,Newman_2013a,RochaI,RochaII}.


Here, we focus on Q-balls~\cite{Coleman1985,Lee1987,Kusenko2001,KimballQball2017}, a well-studied and viable DM candidate.
Formation of Q-balls requires $U(\phi)$ with a global $U(1)$ symmetry. With a $U(\phi)$ satisfying certain conditions~\cite{Coleman1985,Kusenko2001}, small fluctuations in the DM fluid can grow and form stable spherically-symmetric solitons (or clumps)~\cite{KUSENKO199846,Brax2020} with conserved $U(1)$ charge $Q$. As an illustration, we consider ``thin-wall'' Q-balls, where $\phi(r) \approx \phi_{\rm max} e^{i\omega t}$ inside ($r<$ Q-ball radius $R$) and $\phi=0$ outside the clump.
Then, with $V=4\pi R^3/3$ being Q-ball volume, 
 $   Q=\omega\phi^2_{\rm max}V$ and
 its total energy~\cite{Kusenko2001}
$ E=Q^2/(2V\phi^2_{\rm max})+VU(\phi_{\rm max})$. Minimizing $E$ with respect to $V$ yields $R$ and 
\begin{equation}\label{eq:NS_energy}
    E_\mathrm{min}=10\pi\omega^2\phi_{\rm max}^2R^3/3\,.
\end{equation}
Here and below we use natural units.

Let us now turn to a possible non-gravitational interaction between Q-balls and normal matter. Since we are interested in DM-induced variations of $\alpha$, we consider the coupling of $\phi$ to electromagnetism. The simplest phenomenological interaction that preserves the $U(1)$ symmetry reads~\cite{DerPos14,Olive2008}
\begin{equation}\label{Eq:Lagrangian}
    \mathcal{L}_{\mathrm{DM-SM}}\supset\phi^2F_{\mu\nu}^2/\left(4\mathrm{\Lambda}_\gamma^2\right)\,,
\end{equation}
where $\mathrm{\Lambda}_\gamma$ is an unknown characteristic energy scale and $F_{\mu\nu}$ is the Faraday tensor. Inside a clump, $\mathcal{L}_{\mathrm{DM-SM}}$ redefines $\alpha$~\cite{PhysRevLett.115.201301}
\begin{equation}\label{eq:alpha_change}
    \frac{\alpha}{\alpha_0} = \frac{1}{ 1-\phi_{\max}^2/\mathrm{\Lambda}_\gamma^2}\,.
\end{equation}
A constraint on $\alpha/\alpha_0$, translates into bounds on $\mathrm{\Lambda}_\gamma$:
$\Lambda_\gamma=\phi_{\rm max} \, (1-\alpha_0/\alpha)^{-1/2}$. 
As long as $\alpha/\alpha_0\gg 1$, $\Lambda_\gamma \approx \phi_{\rm max}$, a condition largely insensitive to the exact value of $\alpha/\alpha_0$. 

We determine the field amplitude $\phi_{\rm max}$ from the known DM characteristics per standard halo model~\cite{Li_2020} and by requiring that a Q-ball has not enveloped Earth in the past $\mathcal{T}=4$ billion years. To this end, we assume that Q-balls saturate the local DM energy density $\rho_{\rm DM} = n E_\mathrm{min}$, where $n$ is the Q-ball number density and $\rho_{\rm DM} = 0.3(1) \, \mathrm{GeV}/\mathrm{cm}^3$~\cite{Bovy_2012}. The time between consecutive encounters of Q-balls  with Earth is  
\begin{equation}\label{eq:T}
    \mathcal{T}=\frac{1}{v_g}\frac{1}{n\pi R^2}=\frac{10\pi}{3v_g}\frac{\omega^2R\phi^2_{\rm max}}{\rho_{\rm DM}}\,,
\end{equation}
where the average velocity of DM constituents $v_g\approx300\,\mathrm{km/s}$. With fixed $\mathcal{T}=4\, \mathrm{Gy}$, this relates $\phi_{\rm max}$, and, thereby, $\Lambda_\gamma$,  with Q-ball parameters $\omega$ and $R$.

Note that since the DM particles cannot be confined to a region smaller than their Compton wavelength $\lambda_\phi=1/m_\phi$, where $m_\phi$ is the mass of DM field quantum, it is required that $R\gtrsim 1/(2m_\phi)$. An Earth-size Q-ball thus requires $m_\phi\lesssim10^{-14}$ eV. The stability of Q-balls~\cite{Coleman1985} also necessitates $\omega\lesssim m_\phi$. Here, for simplicity, we assume that $\omega\sim m_\phi \sim 1/(2R)$. Our anthropic argument then implies
\begin{equation}\label{eq:Constraint_NS}
    \mathrm{\Lambda}_\gamma\gtrsim3\times{10}^7\left(\frac{\mathcal{T}}{4\,\,\mathrm{Gy}}\frac{R}{R_\oplus}\right)^{1/2}\mathrm{TeV}\,,
\end{equation}
where $R_\oplus\approx6500$ km is Earth’s radius. The anthropic constraint~\eqref{eq:Constraint_NS} is shown in Fig.~\ref{fig:Limits}. It is  about six orders of magnitude more stringent than astrophysical bounds from stellar cooling~\cite{Olive2008}.
\begin{figure}
    \centering
    \includegraphics[scale=0.3]{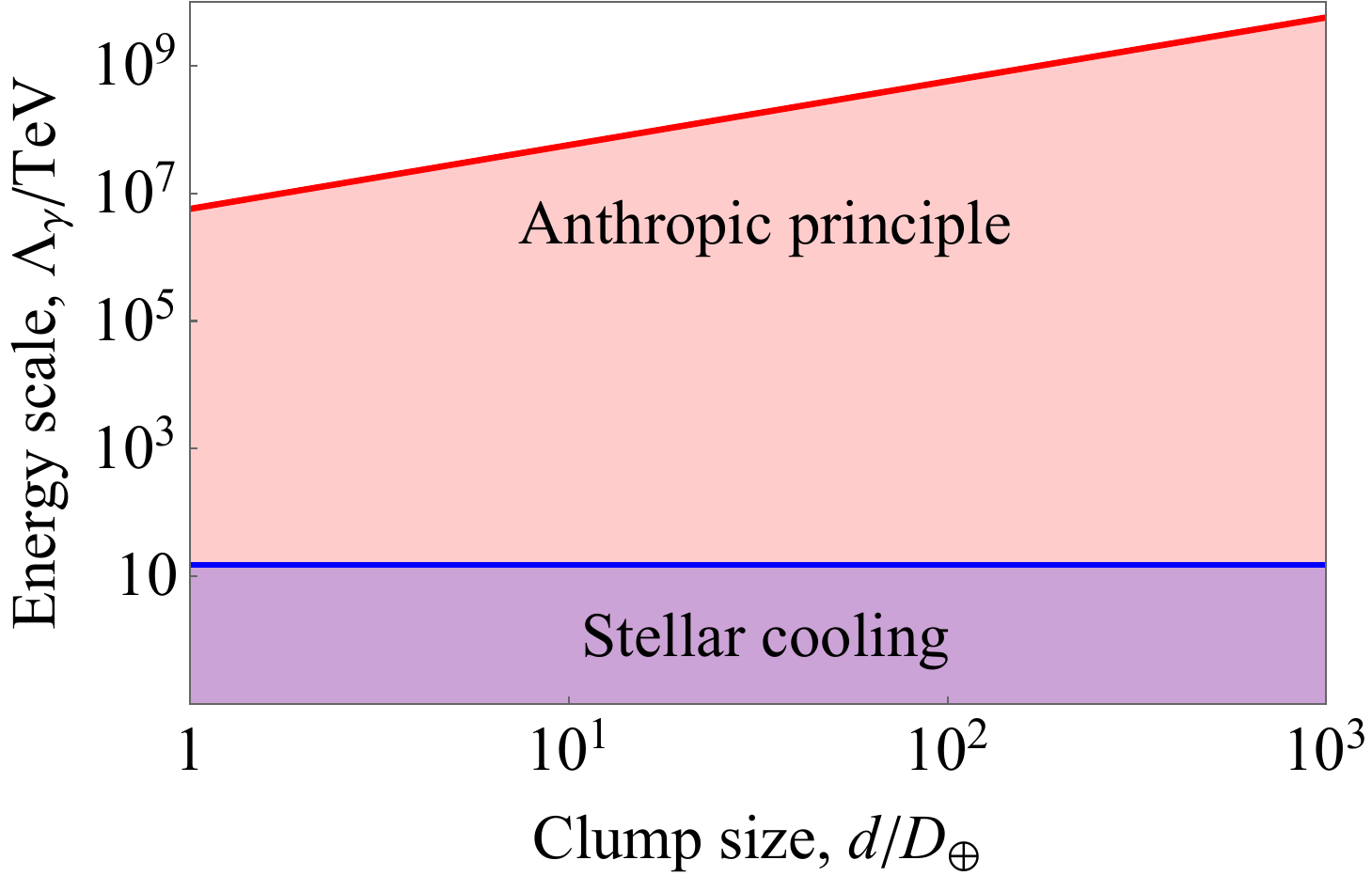}
    \caption{Comparison between anthropic and astrophysical constraints on the energy scale $\Lambda_{\gamma}$ of the interaction $\mathcal{L}_{\rm int}=\phi^2F_{\mu\nu}^2/\left(4\mathrm{\Lambda}_\gamma^2\right)$ between dark matter field and electromagnetism.}
    \label{fig:Limits}
\end{figure}

We end our discussions with the following comments. Firstly, although the limit~\eqref{eq:Constraint_NS} was obtained for the specific case of Q-balls DM, comparable constraints may also be derived for other types of ``clumpy'' DM candidates. See Appendix E, where we derived constraints for topological defect DM.  
Our Fig.~\ref{fig:Limits} limit  seems to be robust across a range of clumpy DM models.

Secondly, it is interesting to ask if DM clumps with large $\phi_{\rm max}$ may form under an extra interaction of the form~\eqref{Eq:Lagrangian}. For Q-balls, the condition for their formation is that the self-interaction potential $U(\phi)$ has a minimum at $\phi=0$. The interaction~\eqref{Eq:Lagrangian} causes the shift $U(\phi)\rightarrow U(\phi)-\phi^2F^2_{\mu\nu}/(4\Lambda^2_\gamma)$. It is clear that the new $U(\phi)$ also has an extremum at $\phi=0$ and retains the U(1) symmetry. For this to be a minimum, it is required that
\begin{equation}\label{cond}
    d^2U(0)/d\phi^2>F^2_{\mu\nu}/(2\Lambda^2_\gamma)\,,
\end{equation}
which may be satisfied for a large class of $U(\phi)$. We note also that once the Q-balls are formed, the extra term $\phi^2F^2_{\mu\nu}/(4\Lambda^2_\gamma)\sim\phi^2\mathcal{E}^2/\Lambda^2_\gamma$, where $\mathcal{E}$ is the electric field, tends to grow smaller due to the Universe expansion (since $\mathcal{E}^2<2 \rho_{\rm EM}<2\rho$, where $\rho_{\rm EM}$ is the radiation energy density and $\rho$ is the total energy density). Thereby, if the condition~\eqref{cond} is satisfied at the Q-ball formation epoch, it will hold at all later times. 

Thirdly, the interaction~\eqref{Eq:Lagrangian} causing large variations of $\alpha$ may cause screening effects preventing the DM clumps from penetrating Earth~\cite{Stadnik2020}. We find that for $\Lambda_\gamma$ satisfying the condition~\eqref{eq:Constraint_NS}, DM clumps are not screened by the atmosphere. They would penetrate up to $\sim 10$ km of water and up to $\sim 3$ km of rocks. These values are large in comparison with the depths at which most life-forms may be found. The screening of DM thus does not affect our anthropic argument. 

Finally, 
a transient exposure to varying FCs would be catastrophic, if it induces detrimental processes that (i) are fast compared to the exposure duration and (ii) are irreversible. As to the duration, a DM clump would sweep through a given molecule over a characteristic timescale of $R/v_g$, which evaluates to $30 \, \mathrm{sec}$ for an Earth-sized clump. On this timescale, we  identify at least two detrimental effects on life building blocks, amino-acids. Both effects lead to nearly identical (due to Eq.~\eqref{eq:alpha_change} properties) anthropic constraints 
on the energy scale $\Lambda_\gamma$, Fig.~\ref{fig:Limits}. Amino-acids contain hydrogen atoms,  among other elements, and we examine effects on hydrogen. First, at $\alpha \sim 100 \alpha_0$, the $1s_{1/2}$ atomic level of hydrogen dives into the  Dirac sea, Fig.~\ref{fig:Dirac}, emitting an electron-positron pair on the $10^{-18} \, \mathrm{sec}$ timescale~\cite{GreRei02QED_book}. Once the pair is emitted, it is absorbed in the surrounding media, making the process irreversible. Second, a similar observation applies to the proton $\beta^+$ decay, $p\rightarrow n+e^+  + \nu_e$. As we show in Appendix F, at $\alpha \gtrsim 5.5 \alpha_0$, the proton lifetime is $\lesssim 30 \, \mathrm{sec}$. The resulting neutron recoils with $\gtrsim$ keV energy; it would rip through the surrounding media causing additional radiation damage. Life is fragile.  
}

\textit{Acknowledgements:} We thank X. Chu, M. Kozlov, A. Kusenko, and M. Pospelov for useful discussions. This work was supported by the National Science Foundation Grants PHY-1912465, PHY-2207546, and CHE-1654547 and the Center for Fundamental Physics at Northwestern University.

\appendix

\section{Invariance of molecular geometry under variation of fundamental constants in the non-relativistic  Born-Oppenheimer approximation}\label{Sec:SM:ScalingInvariance}

\begin{figure}[htb!]
\centering
\includegraphics[width=3in]{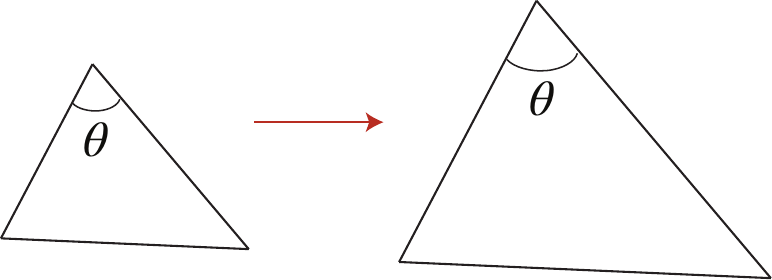}
\caption{
Scaling all the sides of the triangle by the same numerical factor does not affect the value of angle $\theta$ (or of any other angle in the triangle). This example can be generalized to 3D geometry: angles  and thus the molecular geometry are not affected by scaling of all the inter-nuclear distances by the same factor (isotropic scaling transformation).}
\label{Fig:InvarianceOfTheAngles}
\end{figure}

The goal of this section is to prove that the variation of fundamental 
constants cause all the molecular bonds to stretch/dilate by the  very same scaling factor, leaving the angles between molecular bonds unaffected, see Fig.~\ref{Fig:InvarianceOfTheAngles}. This statement holds only with the assumption of (i) the non-relativistic approximation, (ii) infinitely-heavy nuclei (Born-Oppenheimer approximation) and, (iii) point-like spinless nuclei. If any one of these assumptions is violated, molecular bond angles would vary with changing FCs.

We begin by considering an arbitrary molecule containing $N_n$ point-like nuclei and $N_e$ electrons.
Under the enumerated assumptions, the non-relativistic Born-Oppenheimer (NR-BO) Hamiltonian has the form
\begin{align}
   {H}_\mr{NR-BO}&=-\sum_{i}\frac{\hbar^{2}}{2 m_e} \Delta_{{\bf r}_i}
+\frac{1}{2} \sum_{i\neq j} \frac{e^{2}}{\left|\mb{r}_{i}-\mb{r}_{j}\right|}\nonumber\\
&+\frac{1}{2} \sum_{n \neq n^{\prime}} \frac{Z_{n} Z_{n^{\prime}}e^2}{\left|\mb{R}_{n}-\mb{R}_{n^{\prime}}\right|}
-\sum_{i, n} \frac{Z_{n} e^{2}}{\left|\mb{R}_{n}-\mb{r}_{i}\right|}\,.
\end{align}
Here, for clarity, we retained all the fundamental constants (FCs) and we use the Gaussian system of electromagnetic units. We labeled the positions of electrons as $\mathbf{r}_i$ and those of nuclei as $\mathbf{R}_n$, with $Z_n$ being nuclear charges. All the terms in ${H}_\mr{NR-BO}$ have their usual meaning: the kinetic energy of electrons, the electron repulsion, the nuclear repulsion, and the electron-nucleus attraction, respectively. The assumption of infinitely heavy nuclei allows the nuclear kinetic terms to be neglected.

To determine the molecular geometry in the BO approximation, one first solves the time-independent Schr\"{o}dinger equation the positions of the nuclei fixed, 
\begin{equation}
 H_\mr{NR-BO}\left( \mb{r}_{e}| {\mb{R}_{n}}\right) \Psi\left(\mb{r}_{e} | \mb{R}_{n}\right)=E\left(\mb{R}_{n}\right) \Psi\left(\mb{r}_{e} | \mb{R}_{n}\right) \,. \label{Eq:TISE-BO}
 \end{equation}
Here the Hamiltonian $H$ and thus the eigenfunctions $\Psi$ and energies $E$ depend on fundamental constants: $E\left(\mb{R}_{n} | m_e, \hbar, e \right)$.
After the potential surfaces $E\left(\mb{R}_{n} | m_e, \hbar, e \right)$ are obtained as functions of nuclear coordinates, the equilibrium nuclear positions,  $\{\mb{R}_{n}^\mr{eq}\}$, are determined by minimizing the energy
\begin{equation}
\min_{\{\mb{R}_{n} \}} E\left(\mb{R}_{n} | m_e, \hbar, e  \right) \quad \Rightarrow \quad \{\mb{R}_{n}^\mr{eq}\} \,. \label{Eq:FindEqulibrium}
\end{equation}

We would like to explicitly factor out the dependence on FCs from Eq.~(\ref{Eq:TISE-BO}). To this end, we rescale all the coordinates by the same factor $\xi$: $\mb{r}_{i} \rightarrow \xi \bs{\rho}_{i}$,
$\mb{R}_{n} \rightarrow \xi \bs{\rho}_{n}$. Upon substitution into $H_\mr{NR-BO}$, the kinetic energy term is transformed
into $ - \sum_{i}^{N_e}\hbar^{2}\Delta_{\boldsymbol{\rho}_i}/(2 m_e \xi^2) $ and all the electro-static interaction potentials are divided by $\xi$. We may choose $\xi$ so that the dimensionful prefactors in the kinetic and potential energy contributions equal one another, i.e.,  
\begin{equation}\label{Eq:xichoose}
    \frac{\hbar^{2}}{ m_e \xi^2} = \frac{e^2}{\xi} \,.
\end{equation}
This particular choice of $\xi$ enables us to factor out the dependence on the FCs from the Hamiltonian $H_{\rm NR-BO}$. Solving Eq.~\eqref{Eq:xichoose} results in 
\begin{equation}
    \xi = \frac{\hbar^{2}}{ m_e e^2} =a\,,
\end{equation}
where $a$ is the Bohr radius. 

With this choice, the Hamiltonian admits the factorization $H_\mr{NR-BO}= E_h  h(\bs{\rho}_{e}|\bs{\rho}_{n})$, where 
\begin{equation}
    E_h = {\hbar^{2}}/ m_e \xi^2 = {e^2}/{\xi} =m_{e}e^{4}/\hbar^{2}  \label{Eq:Hartree}
\end{equation}
is the atomic unit of energy (Hartree) and $h(\boldsymbol{\rho}_e|\boldsymbol{\rho}_n)$ is a scaled Hamiltonian which does not depend on FCs.
The solution of the eigenvalue equation
\begin{equation}
h(\bs{\rho}_{e}|\bs{\rho}_{n}) \varphi\left(\bs{\rho}_{e}|\bs{\rho}_{n}\right)=
\varepsilon \left(\bs{\rho}_{n} \right) \varphi\left(\bs{\rho}_{e}|\bs{\rho}_{n}\right) \,. \label{Eq:TISE-BO-scaled}
\end{equation}
which is the rescaled version of Eq.~(\ref{Eq:TISE-BO}), does not depend on FCs either. The full energy $E({\bf R}_n)$ is obtained from $\varepsilon(\boldsymbol{\rho}_n)$ via
$$
E\left(\mb{R}_{n} | m_e, \hbar, e  \right) = \frac{m_{e}e^{4}}{\hbar^{2}}  \varepsilon \left(\bs{\rho}_{n} \right)\,,\label{Eq:Scaled_Energy}
$$
which shows that the minimum of $E({\bf R}_n|m_e,h,c)$ is attained at the point 
\begin{equation}
\mb{R}_{n}^\mr{eq} = \frac{\hbar^{2}}{ m_e e^2} \bs{\rho}_{n}^\mr{eq} \,, 
\end{equation}
where $\bs{\rho}_{n}^\mr{eq}$ is obtained by finding the equilibrium positions
\begin{equation}
\min_{\{\bs{\rho}_{n} \}} \varepsilon \left(\bs{\rho}_{n} \right) \quad \Rightarrow \quad \{\bs{\rho}_{n}^\mr{eq}\} \,,
\end{equation}
and is thus FC-independent.

To reiterate, in the non-relativistic Born-Oppenheimer approximation, as 
the FCs are varied from their nominal values, all the equilibrium positions are scaled by  the very same factor,
\begin{equation}
    \mb{R}_{n}^\mr{eq}=\frac{a}{a_{0}} \mb{R}_{n, 0}^\mr{eq} \, . \label{Eq:BohrRadiusScalingNuclear}
\end{equation}
Here and below all the quantities with the subscript $0$  refer to the nominal values. Equation~\eqref{Eq:BohrRadiusScalingNuclear} represents an isotropic scaling of all the coordinates by the same factor, leaving all  molecular bond angles unaffected, see Fig.~\ref{Fig:InvarianceOfTheAngles}.

The fact that the  isotropic scaling does not affect angles in a molecule of arbitrary geometry can be formally proven as follows.
Choose $\{\mb{R}_n\}_{n=1,\ldots,N}$ to be (equilibrium) position vectors of all $N$ nuclei in a molecule.  The angle $\theta_{ab}$ between  a pair of these vectors, $\mb{R}_a$ and $\mb{R}_b$, is given by   
\begin{equation}
    {\theta_{ab}} = \cos^{-1} \left[  \frac{( \mb{R}_a \cdot  \mb{R}_b)}{  |\mb{R}_a| |\mb{R}_b|} \right] \,,
    \label{Eq:SM:Scaling:Angles}
\end{equation}
where we used the conventional definition of scalar products and $|\mb{R}_n|= \sqrt{( \mb{R}_n \cdot  \mb{R}_n) }$ is the length of the vector. 
Should  all the position vectors be scaled by some factor $\lambda$, $\mb{R}_n \rightarrow \lambda \mb{R}_n$, the factors of $\lambda$ in Eq.~(\ref{Eq:SM:Scaling:Angles})  cancel out. Thereby, the angles between molecular bonds are not affected by the isotropic scaling. The entire molecule undergoes isotropic stretching or dilation as  the FCs are varied.

In addition, as follows from our derivation, all the electron coordinates undergo the same isotropic scaling,
\begin{equation}
  \mb{r}_{e} =\frac{a}{a_{0}} \mb{r}_{e, 0} \,.
\end{equation}
In particular, this means that the sizes of electronic clouds and atoms are scaled by the same $a/a_0$ ratio. Another point is that all the energies (both atomic and molecular) are scaled by the atomic unit of energy
\begin{equation}
  E =\frac{E_h}{E_{h,0}} E_0 \,. \label{Eq:BohrRadiusScalingElectronic}
\end{equation}

These observations offer a visualization: as a ``Mr.~Tompkins'' clump, introduced in the main text, sweeps through an atom or a molecule, all the energy levels are gently modulated and the atom or molecule ``breathes'' in accordance with the local values of FCs. This picture  is valid in the regime of sufficiently large and slow  clumps. The clumps need to be sufficiently large, so that there are no gradients of FCs across the extent of the individual atom or molecule. The clumps have to be sufficiently slow, so that the induced perturbation does not cause transitions between molecular or atomic levels. Then the atom or molecule follows the change in FCs adiabatically. 

It is worth emphasizing that our proof heavily relies on the possibility of factoring out all the dependence on FCs in various contributions to the $H_\mr{NR-BO}$ Hamiltonian. 
If we were to the add kinetic 
energies of the nuclei to $H_\mr{NR-BO}$, our coordinate scaling procedure would result in the requirement
\begin{equation}
    \frac{\hbar^{2}}{ m_e \xi^2} = \frac{e^2}{\xi} = \frac{\hbar^{2}}{ M_1 \xi^2} = \cdots = \frac{\hbar^{2}}{ M_N \xi^2} \,,
\end{equation}
where $M_n$ are the nuclear masses. Generically, these equalities can not be satisfied simultaneously  by any choice of the scaling parameter $\xi$.

Our factorization procedure depends on the fact that the Coulomb interactions in the $H_\mr{NR-BO}$  Hamiltonian  exhibited power-law dependence with respect to distances.
If the nuclei have finite size, the Hamiltonian no longer admits simple coordinate scaling. Moreover, introducing nuclear properties (such as finite-size charge distribution or hyperfine interactions with nuclear moments) into the problem brings in another FC, $m_q/\Lambda_\mr{QCD}$,
where $m_q$ is the average mass of up and down quarks and $\Lambda_\mr{QCD}$ is the energy scale of quantum chromo-dynamics. 

Similarly, the Dirac equation does not admit factoring out all the FCs in the Hamiltonian. Indeed, even in the simplest case of the hydrogen atom with an infinitely-heavy point-like nucleus, the Dirac Hamiltonian contains three terms,
\begin{equation}
h_\mr{D}= -i \hbar c \, \bm{\alpha} \cdot \bs{\nabla} +\beta m_e c^{2} - \frac{e^2}{r} \,.
\end{equation}%
Since the $4\times 4$ Dirac matrices $\bm{\alpha}$ and $\beta$ are collections of FC-independent c-numbers, our coordinate scaling procedure results in the requirement
\begin{equation}
  \frac{\hbar c}{\xi} =   m_e c^{2} = \frac{e^2}{\xi}  \,.
\end{equation}
For arbitrary values of FCs ($m_e$, $\hbar$, $e$, and $c$), these equalities cannot be simultaneously satisfied. We conclude that relativity must lead to the breakdown of the isotropic scaling of atomic structure and molecular geometry with varying FCs. Molecular bond angles vary with changing FCs due to relativistic effects.

Since the theory of quantum electrodynamics (QED) is built on the quantization of relativistic fields, field-theoretic effects also lead to the breakdown of the isotropic scaling with varying FCs. This can be easily understood by examining the effects of vacuum polarization by the nucleus~\cite{GreRei02QED_book}. In QED, a nucleus  is immersed into a nuclear-field-polarized cloud of virtual pairs of particles and anti-particles. Vacuum polarization leads to the replacement of the pure Coulomb potential $-Z/r$ of a point-like nucleus by the Uehling potential. The success of our factorization procedure 
depends on the fact that the Coulomb interactions in the NR-BO Hamiltonian  exhibits a power-law dependence with respect to distances. The Uehling potential lacks this  power-law dependence and, thereby, does not admit factoring out FCs in the resulting Hamiltonian.

To recapitulate, the isotropic scaling of molecular geometry preserves angles between molecular bonds in the NR-BO approximation. Molecules ``breathe'' with varying FCs. 

Finally, consider a thought experiment where we compare lengths of two rulers of different chemical composition. Suppose at the nominal values of FCs both rulers have the same lengths. As the FCs change,  both rulers are expanding/contracting by the same factor in the NR-BO approximation. The observer would not be able to tell if FCs have changed. The very same argument applies to transition frequency comparisons: in the NR-BO approximation, all the dependence of electronic 
energies on FCs is governed by the common factor of Hartree energy $m_{e}e^{4}/\hbar^{2}$. 
Corrections to the most basic NR-BO approximation violate this isotropic scaling law: the lengths of two rulers  in our though experiment would  differ for varying FCs. Similarly, the ratios of transition frequencies for two different atoms or molecules would change
with varying FCs.

\section{Critical values of \texorpdfstring{$\alpha$}{}}
\label{Appendix:CriticalAlphas}
As discussed in the main text, the well-known analytical solution of the Dirac equation for hydrogen-like ions with a point-like nucleus is inadequate for determining  the critical values $\acr$ of the electromagnetic fine-structure constant. In this section, we solve this problem using the more realistic finite-sized nucleus model. We also discuss the critical values  of $\alpha$ for multi-electron atoms and the truncation of periodic system at reduced speed of light.

The critical value $\acr$, specific to an atomic energy level, is determined by the requirement that the energy $\varepsilon$ of that level becomes equal to the Dirac sea threshold, $\varepsilon = -2 m_e c^2$ (see the main text). Here and below the rest mass energy $m_ec^2$ is excluded from $\varepsilon$. For some simple models of the nuclear charge density, this problem can be solved analytically by generalizing the solution for a similar problem of determining the critical nuclear charge  for the fixed nominal value of $\alpha$ (see, for example, Refs.~\cite{ZeldovichPopov1972,GreRei02QED_book}).

The analytical solution can be developed for a spherical shell-like nuclear charge density distribution, $\rho_\mr{shell}(r) \propto \delta(r-R)$, where $R$ is the radius of the nuclear charge shell. Inside the nuclear shell, $r<R$, the potential is constant $V(r)=-Z e^2 /R$ and the solution to the Dirac equation is given by the free particle solution regular at the origin. Outside the nuclear shell, the potential is of a pure Coulomb character, $V(r)=-Z e^2 /r$, and  the solution to the Dirac equation is a linear combinations of the regular and irregular relativistic Coulomb wave functions. Setting $\varepsilon = -2 m_e c^2$ and matching the inner and outer solutions at $r=R$, we find, for the $ns_{1/2}$ orbitals, the following transcendental equation for $\acr$ 
\begin{equation}
    \xi \frac{K^\prime_{i \nu}(\xi)}{K_{i \nu}(\xi)} = 2 (\acr Z) \cot{(\acr Z)} \,, \label{Eq:TransAlphaCritical}
\end{equation}
where $\xi = \sqrt{8 Z R/a_0 }$, $K_{i \nu}(\xi)$ is the modified Bessel function of the second kind (also known as the Macdonald function) with index $\nu \equiv 2 \sqrt{(\acr Z)^2 -1}$. $K'_{i \nu}(\xi)$ is the derivative of $K_{i \nu}(\xi)$ with respect to $\xi$. Equation~(\ref{Eq:TransAlphaCritical}) has infinitely many roots, which correspond to increasing values of the principal quantum number $n$. The lowest value of the root determines $\acr$ for the $1s_{1/2}$ orbital.

To solve Eq.~(\ref{Eq:TransAlphaCritical}), we need to specify the nuclear charge shell radius $R$. We make a connection to the more realistic nuclear charge distributions by noticing that for the spherical shell distribution, the root-mean-square (rms) radius $R_\mr{rms}$ is identical to $R$.  For proton, we take the 2018 CODATA~\cite{CODATA-2018} recommended value, $R_\mr{rms}(^1\mr{H}) = 0.8414(19) \, \mathrm{fm}$. For heavier elements, we use an approximation~\cite{GreRei02QED_book} $R \approx 1.6 \, Z^{1/3} \,  \mathrm{fm}$, adequate for our semi-qualitative discussions. From Eq.~(\ref{Eq:TransAlphaCritical}) we find the critical value of $\alpha$ for hydrogen $1s_{1/2}$ to be $\acr\approx 1.04$ or, equivalently, $c^{\star} \approx c_0/143$. 

One may argue that the spherical shell approximation for the nuclear charge distributions used in deriving Eq.~(\ref{Eq:TransAlphaCritical}) is not realistic. To address this question, we solved the Dirac equation for hydrogen numerically using the finite-difference techniques~\cite{WRJBook}; for a uniform nuclear charge distribution we find $\acr\approx 1.042$. This is to be compared to the spherical shell result of $1.040$. A similar exercise for fermium ($Z=100, A = 257, R_\mr{rms}=7.1717 \, \mr{fm}$) shows that the $1s_{1/2}$ value of $\acr$ is $1.20 \times 10^{-2}$ for the spherical shell distribution and $1.18 \times 10^{-2}$ for both the uniform and the Fermi nuclear charge distributions. Thus the accuracy of the spherical shell approximation for the nuclear charge distribution is $\sim 0.1-1\%$, which is adequate for our goals.

We note that the DIRAC19 package internally uses the Gaussian nuclear charge distributions with $R_\mr{rms}$ given by the fitting  formula~\cite{JohSof85} 
\begin{equation}
 R_\mr{rms}=0.836 \, A^{1/3}+0.570\left(\pm 0.05\right)
 \, \mathrm{fm} \,. \label{Eq:RrmsFit}
 \end{equation}
For a given charge $Z$ we use the mass number $A$ for the most abundant isotope. This formula results in the proton $R_\mr{rms}(^1\mr{H}) = 1.406 \, \mathrm{fm}$ which is almost twice as large as the CODATA recommended value, $R_\mr{rms}(^1\mr{H}) = 0.8414(19) \, \mathrm{fm}$.
The simple reason for this discrepancy is that Eq.~(\ref{Eq:RrmsFit}) is a  fit for atomic mass numbers $A>9$, see Ref.~\cite{JohSof85}. If we use the value $R_\mr{rms}(^1\mr{H}) = 1.406 \, \mathrm{fm}$, Eq.~(\ref{Eq:TransAlphaCritical}) results in $\acr(^1\mr{H}) = 1.044$, slightly larger than the value of $1.040$ obtained with the CODATA $R_\mr{rms}(^1\mr{H})$.

\begin{figure}[hbt!]
\centering
\includegraphics[width=3in]{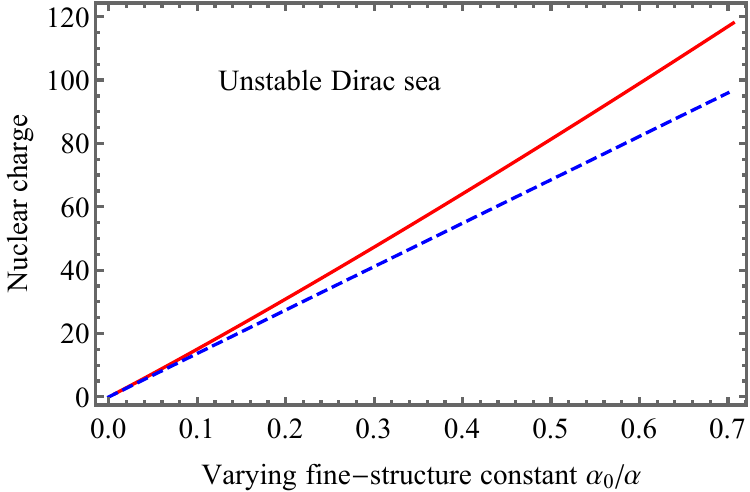}
\caption{Phase diagram of periodic system of elements as a function of varying fine-structure constant. 
Red curve are the results for critical values of $\alpha_0/\alpha^\star$ as a function of nuclear charge $Z$ for finite-size nuclei. Blue curve is the same dependence but for point-like nuclei. To borrow an analogy from condensed matter physics, $\alpha$ (or $c$) is an order parameter that governs phase transitions.  
}
\label{Fig:PhaseDiagram}
\end{figure}

The results of our calculations for $\acr$ as a function of $Z$ are shown in Fig.~\ref{Fig:PhaseDiagram}. In this plot, the red curve represents the relationship between $\alpha_0/\alpha^\star$ and the nuclear charge $Z$ of a finite-size nucleus with a Fermi charge distribution. The blue curve represents the same dependence but for a point-like nucleus, where $\acr =1/Z$. This parameter space can be interpreted as a phase diagram: any point $(\alpha_0/\alpha,Z)$ lying above the red curve corresponds to unstable Dirac sea, where the $1s_{1/2}$ orbital is embedded into the Dirac sea continuum. 

\begin{figure}[htb!]
\centering
\includegraphics[width=3in]{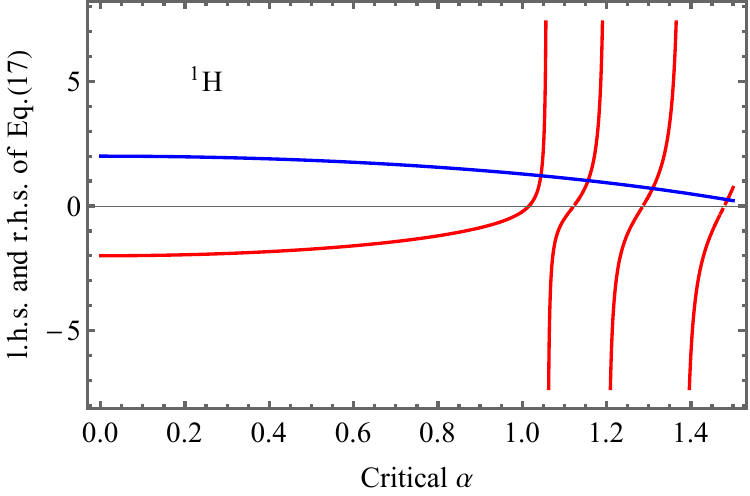}
\caption{Graphical determination of critical values of fine-structure constant for hydrogen ($Z=1, R= 0.8414 \, \mathrm{fm}$). The r.h.s and the l.h.s of transcendental Eq.~(\ref{Eq:TransAlphaCritical})  are drawn as  red and blue curves, respectively. The values of $\alpha$ at the intersection of two curves are critical values $\acr$ of $\alpha$. The lowest $\acr$  is the critical value for $1s_{1/2}$, next lowest $\acr$ is for $2s_{1/2}$ and so on.
 }
\label{Fig:GraphSolCriticalAlpha}
\end{figure}

The finite-size nuclei critical curve exhibits a nearly linear dependence with a fit,
\begin{equation}
    Z_\mr{max} \approx 168 \alpha_0/\alpha\,, \label{Eq:Zmax:LinearFit}
\end{equation}
or, equivalently,
\begin{equation}
   \alpha_0/\alpha^\star\approx Z/168 \,. \label{Eq:cStar:LinearFit}
\end{equation}
\begin{equation}
    \alpha < 1.23/Z
\end{equation}
The linearity of these equations can be understood by examining the
graphical solution of the transcendental Eq.~(\ref{Eq:TransAlphaCritical}), see  Fig.~\ref{Fig:GraphSolCriticalAlpha} for hydrogen; plots for heavier elements are similar. 
Even without solving the equation~(\ref{Eq:TransAlphaCritical}), it is apparent that the  critical value of $\alpha$ for the $1s_{1/2}$ orbital occurs in the vicinity of the first zero of Macdonald function $K_{i \nu}(\xi)$, where the l.h.s. approaches the  vertical asymptote. The first zero of $K_{i \nu}(\xi)$ is given
by~ $\ln \xi \approx - \pi/\nu + \ln 2 - \gamma_\mr{Euler}$,  where $\gamma_\mr{Euler}= 0.5772156649\ldots$ is the Euler constant~\cite{Ferreira2008}. This leads to an analytical estimate
\begin{equation}
    \acr  \lesssim  \frac{1}{Z} \left( 1
     + \frac{\pi^2}{8}  \frac{1}{ \left(\gamma_\mr{Euler} +\frac{1}{2} \ln(\sqrt{2} Z R/a_0 )\right)^2} \right) \,. \label{Eq:ApproximateAlphaCritical}
\end{equation}

The leading $1/{Z}$ term in Eq.~\eqref{Eq:ApproximateAlphaCritical} can be recognized as the critical value for the point-like nucleus (see the main text).
We use the smaller sign ($\lesssim$) because the true value of $\acr$ lies below this asymptotic estimate,
see Fig.~\ref{Fig:GraphSolCriticalAlpha}.  The second term is due to the finite size of the nucleus, with the nuclear radius $R\propto Z^{1/3}$.
The fractional contribution of this second term has a weak logarithmic dependence on the nuclear charge, $\ln(Z^{4/3})$, explaining the nearly linear dependence of the maximum allowed nuclear charge in Fig.~\ref{Fig:PhaseDiagram}. In the approximate Eq.~(\ref{Eq:ApproximateAlphaCritical}) we also restored the Bohr radius $a_0$, showing that the dominant dependence is the ratio of the nuclear radius $R$ to the characteristic size of the atomic orbital $a_0/Z$. The approximation~(\ref{Eq:ApproximateAlphaCritical}) tends to overestimate $\acr$. Its relative accuracy ranges from 2\% for hydrogen to 50\% for Fermium ($Z=100$) as follows from a comparison with our numerical results. It is worth noting also that QED corrections to the $1s_{1/2}$ energy (vacuum polarization and self-energy) 
tend to cancel~\cite{GreRei02QED_book}, leaving the critical value $\alpha^\star$ largely unaffected by these corrections. 

For  multi-electron systems, the stability of an atom with respect to varying FCs requires further qualifiers. As discussed in the main text for hydrogen-like ions, for $\alpha>\alpha^\star$, the bound $1s_{1/2}$ level becomes embedded into the Dirac sea continuum, and, as such, becomes unstable, similar to  Fano resonances in chemical physics~\cite{Fri04}. An electron-positron pair is emitted spontaneously and the vacuum becomes electrically charged. For low-lying energy states of  multi-electron atoms, however, the $1s_{1/2}$ orbital is fully occupied. Then a Dirac sea electron can not transition into the fully occupied $1s_{1/2}$ orbital due to the Pauli exclusion principle~\cite{ZeldovichPopov1972}.  Yet, because the rest-mass energy gap is lowered with increased $\alpha$, ambient photons can promote Dirac sea electrons into unoccupied orbitals, i.e., Dirac sea becomes unstable with respect to the interaction with ambient photons. 

As for the critical values of multi-electron systems, we computed the Dirac-Hartree-Fock (DHF) energies of the $1s_{1/2}$ orbitals in several atoms as a function of $\alpha$. We find that the  hydrogen-like ion result~(\ref{Eq:Zmax:LinearFit}) for $\alpha^\star$ remains an excellent approximation. Indeed, the energies of the deeply-bound $1s_{1/2}$ orbitals in atoms and molecules are strongly dominated by the interaction with the nucleus with a small corrections from the interaction with other electrons. 

\section{Results for many-electron atoms}

In this section, we discuss the effects of increasing $\alpha$ on the spectrum of a many-electron atom. We will focus on neon. Results for other second-row elements may be found in Ref.~\cite{Dergachev2022}.

All the $\emph{ab initio}$ electronic structure calculations were performed in the DIRAC19 relativistic quantum chemistry code~\cite{Saue2020}. The Dirac-Hartree-Fock (DHF) wave function was first obtained as a reference for subsequent electron correlation calculations. For all the atoms, the DHF wave function was of the open-shell character, including neon. This was due to inclusion of the $3s_{1/2}$ orbital in the valence space to assess an effect of the relativistic stabilization of this orbital at larger $\alpha$. The open-shell calculations were implemented in the average-of-configuration (AOC) framework~\cite{Saue1999}, where the wave function and energy were optimized for a limited set of open-shell states. The Slater determinants, arising from the active space, that is all possible distributions of valence electrons on selected active orbitals, formed the N-particle basis for the AOC open-shell wave function. The active space included the valence $2p$ electrons and $2p$ and $3s_{1/2}$ orbitals for all atoms. For example, for nitrogen, the active space included three valence $2p$ electrons and four orbitals ($2p_{x}$, $2p_{y}$, $2p_{z}$, and $3s$) in the orbital notation, or eight spin-orbitals ($2p_{1/2}$, $2p_{3/2}$, $3s_{1/2}$) in the spin-orbital notation. The Kramer Restricted Configuration Interaction method with single and double excitations (KRCISD) was used to account for the electron correlation and quasi-degeneracy of the $2p$ orbitals at nominal $\alpha$. The same method was used throughout different $\alpha$ for consistency. The KRCISD calculations were performed on top of the AOC DHF wave function excluding the complete active space self-consistent field (CASSCF) step conventionally included in the electron correlation calculations. This was due to the current  state-specific single-state implementation of the CASSCF method. The state-specific implementation does not allow for averaging the wave function over multiple electronic states thus providing a balanced description of these states. In addition, the single-state framework allows optimization of a wave function only for the lowest state of the given symmetry, excluding higher electronic states. On the other hand, the AOC DHF wave function was balanced over determinants comprising electrons on $2p$ and $3s$ orbitals, whose population determines the lowest excited states of the second-row atoms, and therefore, was used as a reference.

At the nominal $\alpha$, the atomic states are labeled in the conventional L-S (Russell-Saunders) coupling scheme: $^{2S+1}\!L_J$. At increased $\alpha$, however, amplified relativity leads to the breakdown of the LS coupling scheme, as only the total angular momentum $\mathbf{J}= \mathbf{L} + \mathbf{S}$ is conserved~\cite{WRJBook}. 
Thereby, we label the states as $J^{\mathrm{\pi}}$, where $J$ is the value of the total angular momentum and $\mathrm{\pi}$ is the parity of the state. If there are multiple states of the same $J^{\mathrm{\pi}}$ symmetry, we distinguish them by appending their sequential number $n$: $J^{\mathrm{\pi}}(n)$, where the states are enumerated in the order of increasing energy. In our notation for electronic configurations, for brevity, we suppress the $1s_{1/2}$ and  $2s_{1/2}$ shells, as these remain always doubly occupied for our considered low-lying energy states.

Since the typical distance of an electron from the nucleus decreases with increasing $\alpha$ due to relativistic contraction, the basis set used in our calculation needed to be calibrated to accurately describe the electronic density near the nucleus at increased $\alpha$. The calibrating procedure was carried out by considering the hydrogen-like ions of Ne as follows. We gradually increased $\alpha$ until the $1s_{1/2}$ ground state of ${\rm Ne}^+$ dived into the Dirac sea. The size of the basis set and the largest exponents were chosen to match the critical values of $\alpha^{\star}$ obtained using such a basis set with that predicted by solving the transcendental Eq.~(\ref{Eq:TransAlphaCritical}). Additionally, the validity of the basis set was verified by comparing the energy level orderings they generated with those predicted by the finite-difference solution of the Dirac equation. To obtain the correct energy ordering, it required to augment standard basis sets with additional $p$ basis functions.


For Ne, the eleven $p$ basis functions in the original unc-aug-cc-pV6Z basis set were augmented to a total of nineteen. The exponents of the new functions were obtained by subsequently multiplying the largest $p$ exponent by 3. For simplicity, the same exponents were used for the $s$ basis functions. This procedure was carried out until a match with the solution to the transcendental Eq.~(\ref{Eq:TransAlphaCritical}) was obtained, while maintaining the correct ordering of the energy levels. The resulting modified basis thus included nineteen $s$ and nineteen $p$ basis functions with the largest exponent being 1.4 $\times$ 10$^{9}$. Basis functions of higher angular momenta were left unchanged. With such a modified basis set, the critical $\alpha^\star$ value obtained using the DIRAC19 program matched that predicted by Eq.~(\ref{Eq:TransAlphaCritical}), namely, $\alpha^{\star}_{\rm Ne}\approx 14.8\alpha_0$.

Note that to avoid the collapse of many-electron wave functions into the Dirac sea~\cite{BroRav51}, the so-called ``no-pair'' Hamiltonian~\cite{Sap98,RelQC2010_book} was used in fully relativistic electronic structure calculations
\begin{equation}
H_\mr{no-pair}=\sum_{i}  h_\mr{D}(i)   +\frac{1}{2}\sum_{i\neq j} \Lambda_{++}\frac{1}{r_{ij}}\Lambda_{++} \,.
\label{Eq:no-pair-Hamiltonian}
\end{equation}
Here the first term is a sum of the Dirac Hamiltonian $h_\mr{D}(i)$ describing an electron $i$ moving in the potential of a finite-size nucleus and the second term describes the Coulomb repulsion between the electrons. Notice that the $e-e$ interaction is sandwiched between projection operators $\Lambda_{++}$ which exclude states from the Dirac sea continuum of $h_\mr{D}$.

We have described the methods with which we computed the low-lying energy states of Ne at different values of $\alpha$. Below, we present the results of our calculation. We find that as $\alpha$ is increased, the energies of the excited states of Ne exhibit various interesting features. Relative to the energy of the ``nominal'' ground state (the ground state at nominal $\alpha$), an excited state energy may rise or fall in the regime $\alpha\sim 10\alpha_0$, leading to several crossings of levels. However, as $\alpha$ is increased further, all excited states eventually stabilize with respect to the nominal ground state. Even more remarkably, as $\alpha$ nears $\alpha^\star$, the energies of some excited states decrease so much that these states become the ``new'' ground states themselves. 

We show below that the electron-configuration picture is sufficient for the qualitatively explanation of these phenomena. The lynch-pins of this exposition are the facts that as $\alpha$ increases, (i) the $2p_{1/2}$ energy decreases, (ii) the $2p_{3/2}$ energy increases, (iii) the $3p_{1/2}$ energy falls, but with a slower rate than that of $2p_{1/2}$, (iv) the $3p_{3/2}$ energy rises but with a slower rate than that of $2p_{3/2}$, and (v) all $s_{1/2}$ energies fall. It is worth stressing that the rise in the energy of $2p_{3/2}$ (and similarly of all other $p_{3/2}$) orbitals is only present in multi-electron atoms. In a hydrogen-like atom, although the fine-structure contribution to a $2p_{3/2}$ energy increases with increasing $\alpha$, the gross-structure contribution decreases (it is a negative quantity whose magnitude gets larger), leading to an overall decline of the $2p_{3/2}$ energy. In a multi-electron atom, however, the contractions of the inner $1s_{1/2}$, $2s_{1/2}$, and $2p_{1/2}$ orbitals with increasing $\alpha$ leads to more effective screening of the nuclear charge, thus reducing the magnitude of the gross-structure contribution to the $2p_{3/2}$ energy. This in turn means that the $2p_{3/2}$ energy increases with increasing $\alpha$.

\begin{figure}
\centering
\includegraphics[width=\columnwidth]{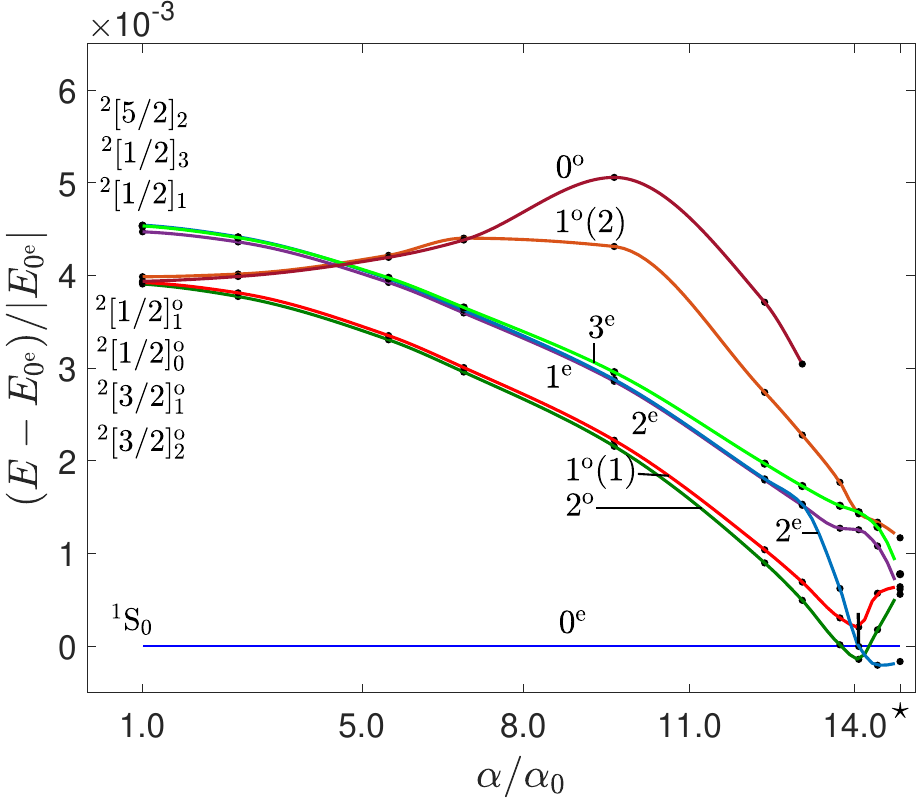}
\caption{Energy spectrum of neon atom as a function of $\alpha$. The nominal ground state $0^e (^1\!S_0)$ is used as a reference (blue horizontal line). At the nominal $\alpha$, the levels are labeled using the conventional L-S coupling scheme~\cite{NIST_ASD}. At larger values of $\alpha$, the levels are labeled as $J^\pi$, where $J$ is the total angular momentum and $\pi$ is the parity of the state. There is a substantial reshuffling of sequence of energies near the critical value of $\alpha\sim\alpha^{\star} \approx 14.8\alpha_0$.
In this regime, the ground state of neon atom becomes the open-shell $2^e$ state.}
\label{Fig:neon}
\end{figure}

With this general picture in mind, let us discuss in details the changes in Ne atomic spectrum as $\alpha$ increases. At the  nominal $\alpha=\alpha_{0}$, the closed-shell $2p^{6}$ ground electron configuration $2p^{6}$ of neon results in a single atomic state with even parity, $^1\mathrm{S_0}$. The $2p^53s$ first excited configuration gives rise to four excited states with odd parity: $^2[3/2]^\mathrm{o}_2$, $^2[3/2]^\mathrm{o}_1$, $^2[1/2]^\mathrm{o}_0$, and $^2[1/2]^\mathrm{o}_1$. The second excited configuration, $2p^53p$, generates a manifold of atomic states with even parity, among which we consider the three lowest states, $^2[1/2]_1$, $^2[5/2]_3$, and $^2[5/2]_2$. We therefore include a total of eight atomic states of neon in our discussion, see Fig.~\ref{Fig:neon} in the main text. At larger $\alpha$, these states are labeled, in the order they are introduced above, as $0^\mathrm{e}$, $2^\mathrm{o}$, $1^\mathrm{o}$(1), $0^\mathrm{o}$, $1^\mathrm{o}$(2), $1^\mathrm{e}$, $3^\mathrm{e}$, and $2^\mathrm{e}$, respectively.

Throughout this discussion, the nominal ground state $0^{\rm e}$, which corresponds to $^1\mathrm{S_0}$ at $\alpha=\alpha_{0}$, is used as a reference; and all state energies are reported as the ratio $(E-E_{0^{\rm e}})/|E_{0^{\rm e}}|$. This choice does not imply, however, that the state $0^{\rm e}$ remains unaffected as $\alpha$ increases. In the range $\alpha_0\leq \alpha\lesssim 13\alpha_0$, the $0^\mathrm{e}$ state retains $0^{\rm e}$ its closed-shell configuration of $2p^2_{1/2}2p^4_{3/2}$. However, at $\alpha\approx 13\alpha_0$, the $3s_{1/2}$ and $2p_{3/2}$ energies become close enough so that $2p^2_{1/2}2p^2_{3/2}3s^2_{1/2}$ emerges as an appreciable contribution in the CI expansion of $0^\mathrm{e}$. Note that two electrons are transferred from $2p_{3/2}$ to $3s_{1/2}$ to preserve total angular momentum and parity. The appearance of the open-shell configuration $2p^2_{1/2}2p^2_{3/2}3s^2_{1/2}$ in the CI expansion of the nominal ground state indicates that at $\alpha\gtrsim 14\alpha_0$, neon is no longer chemically inert.

The fact that increasing $\alpha$ has the effect of ``activating'' the naturally inert neon may also be understood by considering the state $2^{\rm o}$, which corresponds to $^2[3/2]_2^{\rm o}$ at $\alpha=\alpha_0$. The main configuration for $2^{\rm o}$ in the regime $\alpha_0\leq \alpha\lesssim 13\alpha_0$ is $2p^2_{1/2}2p^3_{3/2}3s_{1/2}$ containing the orbital $3s_{1/2}$ which has the effect of destabilizing $2^{\rm o}$. As discussed in the previous paragraph, in the vicinity of $\alpha\approx 13\alpha_0$, the ground state $0^\mathrm{e}$ acquires the component $2p^2_{1/2}2p^2_{3/2}3s^2_{1/2}$. Since the $3s_{1/2}$ orbital still lies above $2p_{3/2}$ in this regime, the $2^{\rm o}$ energy falls below that of $0^\mathrm{e}$ and $2^{\rm o}$ briefly becomes the ``new'' ground state of neon. However, as $\alpha$ nears $14\alpha_0$, the $2p_{3/2}$ and $3s_{1/2}$ orbitals cross, raising $2^{\rm o}$ back above $0^\mathrm{e}$. 

The state $1^{\rm o}(1)$ (nominally $^2[3/2]_1^{\rm o}$) displays a dependence on varying $\alpha$ similar to that of $2^{\rm o}$. In the regime $\alpha_0\leq \alpha\lesssim 13\alpha_0$, its CI expansion is dominated by $2p^2_{1/2}2p^3_{3/2}3s_{1/2}$ and $2p_{1/2}2p^4_{3/2}3s_{1/2}$ which stabilize its energy relative to the ground state. However, due to the second configuration where $2p_{3/2}$ is quadruply occupied, the energy of $1^{\rm o}(1)$ does not lower as dramatically as that of $2^{\rm o}$. In particular, the $1^{\rm o}(1)$ energy never falls below the ground state energy. As $\alpha$ approaches then passes $14\alpha_0$, $3s_{1/2}$ crosses below $2p_{3/2}$ and $1^{\rm o}(1)$ destabilizes relative to $0^{\rm e}$. The rate of destabilization is reduced as $\alpha$ approaches $\alpha^\star$ due to the appearance of the configuration $2p^2_{1/2}2p_{3/2}3s^2_{1/2}4s_{1/2}$ in the expansion of $1^{\rm o}(1)$.

The last odd-parity state included in our discussion is $0^{\rm o}$ (nominally $^2[1/2]^{\rm o}_0$) which, for $\alpha_0\leq \alpha\lesssim 10\alpha_0$, comprises mainly of $2p_{1/2}2p^4_{3/2}3s_{1/2}$ which, similarly to the case of $1^{\rm o}(2)$, causes $0^{\rm o}$ to destabilize relative to the nominal ground state. However, unlike $1^{\rm o}(2)$, the absence of the configuration $1^{\rm o}(2)$ in the expansion of $0^{\rm o}$ means that its energy continues to rise until $\alpha\approx 10\alpha_0$, where an avoided crossing with higher (nominally) $^2[1/2]^{\rm o}_0$ state (not shown) changes the configuration of $0^{\rm o}$ to $2p^2_{1/2}2p^3_{3/2}3d_{5/2}$. Since the $3d_{5/2}$ orbital destabilizes with increasing $\alpha$ at a much slower rate than $2p_{3/2}$, the $0^{\rm o}$ energy experiences a steep downturn relative to the ground state energy for $\alpha\gtrsim 10\alpha_0$.

Next, we consider the even-parity excited states $1^{\rm e}$ and $2^{\rm e}$, which correspond to $^2[1/2]_1$ and $^2[5/2]_2$ at $\alpha=\alpha_0$. These states both start out at as combinations of $2p^2_{1/2}2p^3_{3/2}3p_{1/2}$ and $2p^2_{1/2}2p^3_{3/2}3p_{3/2}$. As $\alpha$ increases, the $3p_{1/2}$ energy decreases and the $3p_{3/2}$ increases at a much slower rate than that of $2p_{1/2}$. As a result, the states $1^{\rm e}$ and $2^{\rm e}$ generally destabilize with respect to the nominal ground state $0^{\rm e}$. However, the stabilization pattern of $2^{\rm e}$ display a peculiar feature. At $\alpha\approx 13\alpha_0$, the configuration $2p^2_{1/2}2p^3_{3/2}3p_{3/2}$ is replaced by $2p^2_{1/2}2p^2_{3/2}3s^2_{1/2}$ from the CI expansion of $1^{\rm e}$ and $2^{\rm e}$. As a result, the decrease of the $2^{\rm e}$ energy becomes even more precipitous and at $\alpha\approx 14\alpha_0$, it becomes less than the $0^{\rm e}$ energy, thus making $2^{\rm e}$ the new ground state of neon. 

Finally, we consider the state $3^{\rm e}$, which is labeled $^2[5/2]_3$ at $\alpha_0$. The dominant configuration in this state remains $2p^2_{1/2}2p^3_{3/2}3p_{3/2}$ for all $\alpha_0\leq\alpha\lesssim\alpha^\star$. As a result, it destabilizes continuously relative to the ground state, albeit not rapidly enough for it to cross the nominal ground state anywhere in the interval $\alpha_0\leq\alpha\lesssim\alpha^\star$.

\section{Results for molecules. Example of a water molecule}

\begin{figure}[h!]
\centering
\includegraphics[width=3in]{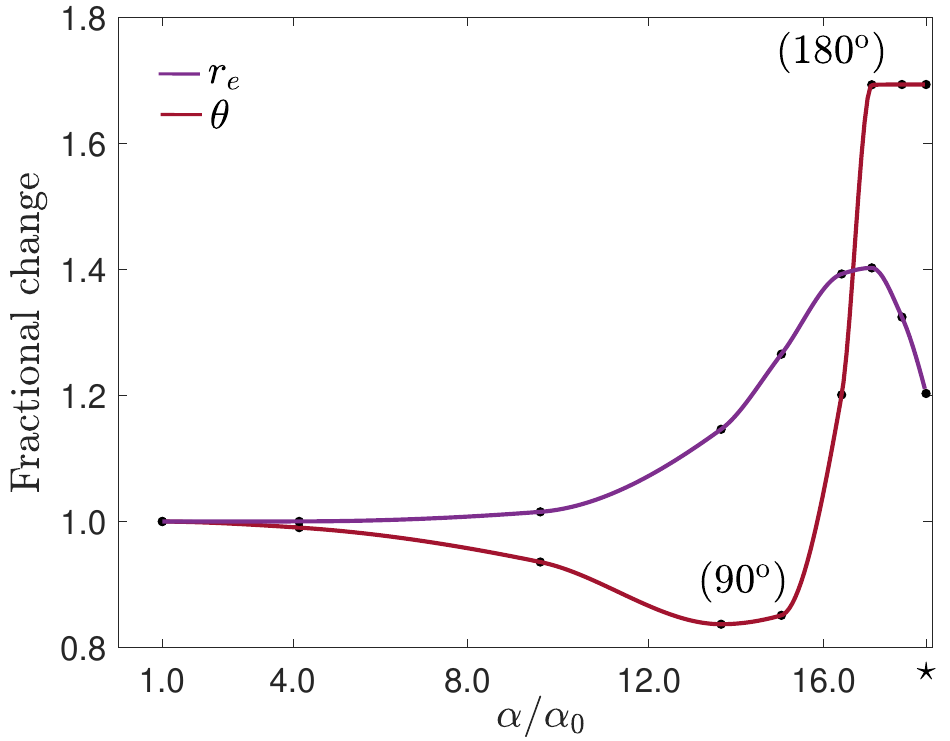}
\caption{Fractional changes of the bond distance $r$ and the bond angle $\theta$ in the water molecule at different values of $\alpha$. The critical value of $\alpha$ for oxygen is $\alpha^\star_\mathrm{O}= 18.4\alpha_0$.}
\label{Fig:Water_structure}
\end{figure}

The changes in the molecular geometry of water can also  understood in the context of the textbook valence-shell electron-pair repulsion (VSEPR) model 
\begin{itemize}
    \item At nominal $\alpha$, the valence $2s$ and $2p$ atomic orbitals of oxygen mix to form four equivalent hybrid orbitals ($sp^3$ hybridization, or $2s_{1/2}2p_{1/2}2p_{3/2}$ in relativistic notation). Two of the hybrid orbitals overlap with the hydrogen $1s$ orbitals, and the remaining two hold the two lone electron pairs. The repulsion between the four electron pairs on the hybrid orbitals leads to the slightly distorted tetrahedral arrangement, corresponding to the bond angle of ${104.5}^\mathrm{o}$.
    
    \item At the intermediate value $\alpha\approx14\alpha_0$, the stabilization of the $2s_{1/2}$ and $2p_{1/2}$ orbitals in oxygen breaks down the $sp^3$ hybridization. This results in an energetically isolated $2p_{3/2}$ orbital, which can accommodate up to four electrons forming the oxygen-hydrogen bonds. This stabilizes the molecular structure with the ${90}^\mathrm{o}$ bond angle~\cite{Dubillard2006}. 
    
	\item At larger $\alpha\approx18\alpha_0$, the lowering of the $3s_{1/2}$ energy and the raising of the $2p_{3/2}$ energy in oxygen leads to these two orbitals becoming quasi-degenerate. This induces the $ps$ hybridization between the half-filled $2p_{3/2}$ and the $3s_{1/2}$ shells~\cite{Dubillard2006}, resulting in the linear molecular geometry.
\end{itemize}
The changes of the bond length and bond angle in the water molecule as functions of $\alpha$ are shown in Fig.~\ref{Fig:Water_structure}.
	

In the relativistic picture, the states of atoms and molecules are described by four-component Dirac spinors $\psi=(\psi^{\alpha}_{\mathrm{L}},\psi^{\beta}_{\mathrm{L}},\psi^{\alpha}_{\mathrm{S}},\psi^{\beta}_{\mathrm{S}})^\mathrm{T}$, where $\mathrm{L}$ and $\mathrm{S}$ indicate the large and small components, respectively, while $\alpha$ and $\beta$ describe the spin degrees of freedom. The spinor components are, in general, complex numbers so a general collection of four such components has eight degrees of freedom. However, since the spatial and spin degrees of freedom are coupled, the symmetry of the Dirac spinors is described by the double groups, where the total spinor transforms under fermion irreducible representations spanned by half-integer spin functions~\cite{Saue1999}. Furthermore, the real and imaginary parts of each spinor component are spanned by boson irreducible representations, which are irreducible representations of conventional single point groups. Therefore, each spinor component can be described by scalar functions, or orbitals. 

The symmetry of the water molecule is described by the $C_{2v}$ double group. For example, exploiting the symmetry of the Dirac Hamiltonian, it can be shown that in the $C_{2v}$ double group, the real and imaginary parts of the large component transform under the ($a_{1}$,$a_{2}$) and ($b_{1}$,$b_{2}$) boson irreducible representations for $\psi^{\alpha}_{\mathrm{L}}$ and $\psi^{\beta}_{\mathrm{L}}$, correspondingly~\cite{DIRAC21}. At nominal $\alpha$, molecular orbitals of water are spanned only by a real or imaginary part of a single component, neglecting vanishing contribution from other components. Therefore, these orbitals are described by a single irreducible representation, in compliance with results from non-relativistic calculations. At increased $\alpha$, however, the molecular orbitals are spanned by multiple real and imaginary parts of the spinor components and one cane no longer assign a single irreducible representation to molecular orbitals. For this reason, the symmetry labels in the MO diagrams of water are presented only at the nominal speed of light, but not at larger $\alpha$. 

In the MO diagram of water (Fig.~2 A-C in the main text), the $\sigma$ and $\sigma^{*}$ linear combinations of $1s$ orbitals of two hydrogen atoms have the $a_1$ and $b_2$ symmetries. For oxygen at nominal $\alpha$, the atomic orbitals $2s_{1/2}$ and $3s_{1/2}$ have the symmetry of $a_1$ whereas the $2p$ orbitals have the symmetries of $a_1$, $b_2$, and $b_1$. In atomic calculations of oxygen, the $3s_{1/2}$ spinor was included in the average-of-configuration Dirac-Hartree-Fock method to assess the effect of stabilizing higher lying spinors on molecular bonding. For clearer comparison of diagrams, we keep the energy unit constant and equal to that at nominal $\alpha$. 

\begin{figure}[h!]
\centering
\includegraphics[width=3.2in]{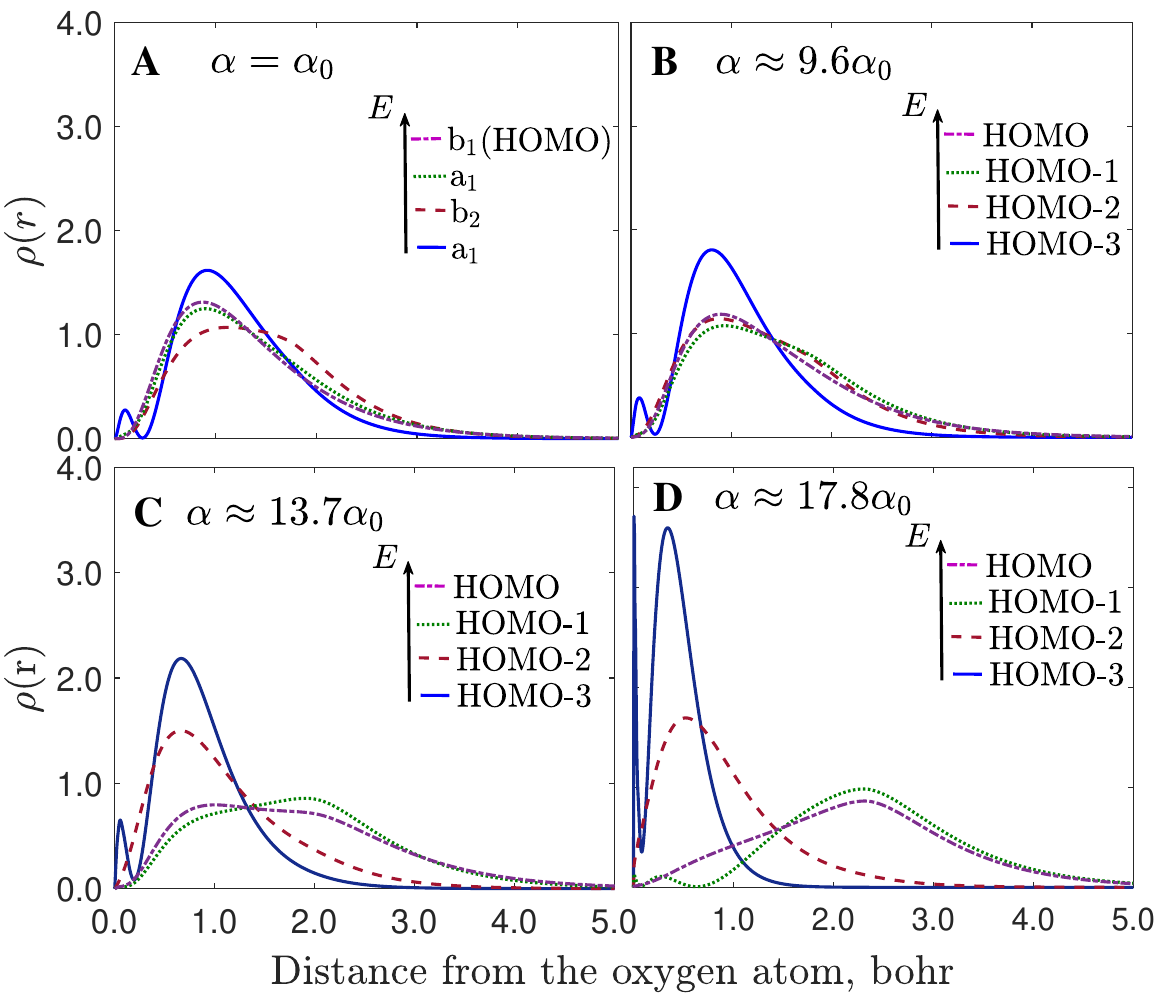}
\caption{Radial density distributions for four occupied molecular orbitals of water at different values of $\alpha$. The oxygen atom is placed at the coordinate origin. The molecular orbitals are labeled in the order of increasing energy. The dark blue solid curve shows the distribution for the lowest occupied molecular orbital, HOMO-3 (HOMO stands for the highest occupied molecular orbital); the dark red dashed curve - HOMO-2; the dark green dotted curve - HOMO-1; and the magenta dot-dashed curve - HOMO. For better resolution, the distance from the oxygen atom is given in the units of the unscaled nominal Bohr radius.}
\label{Fig:Water_RDD}
\end{figure}

To better demonstrate the changes in the electronic structure of water at increased $\alpha$, we calculated the radial density distribution in DIRAC19~\cite{Saue1999,Saue2020} for each molecular orbital of water as
\begin{equation}
 \rho(r) = \int_{0}^{2\pi} \int_{0}^{\pi} \rho(\bm{r})r^2\sin{\theta} \,d\theta\,d\phi\,,
\end{equation}
where $\rho({\bf r})\equiv|\Psi({\bf r})|^2$ is the electron density and $\Psi({\bf r})$ is the MO wave function (see Fig.~\ref{Fig:Water_RDD}).

\begin{figure}[h!]
\centering
\includegraphics[width=3in]{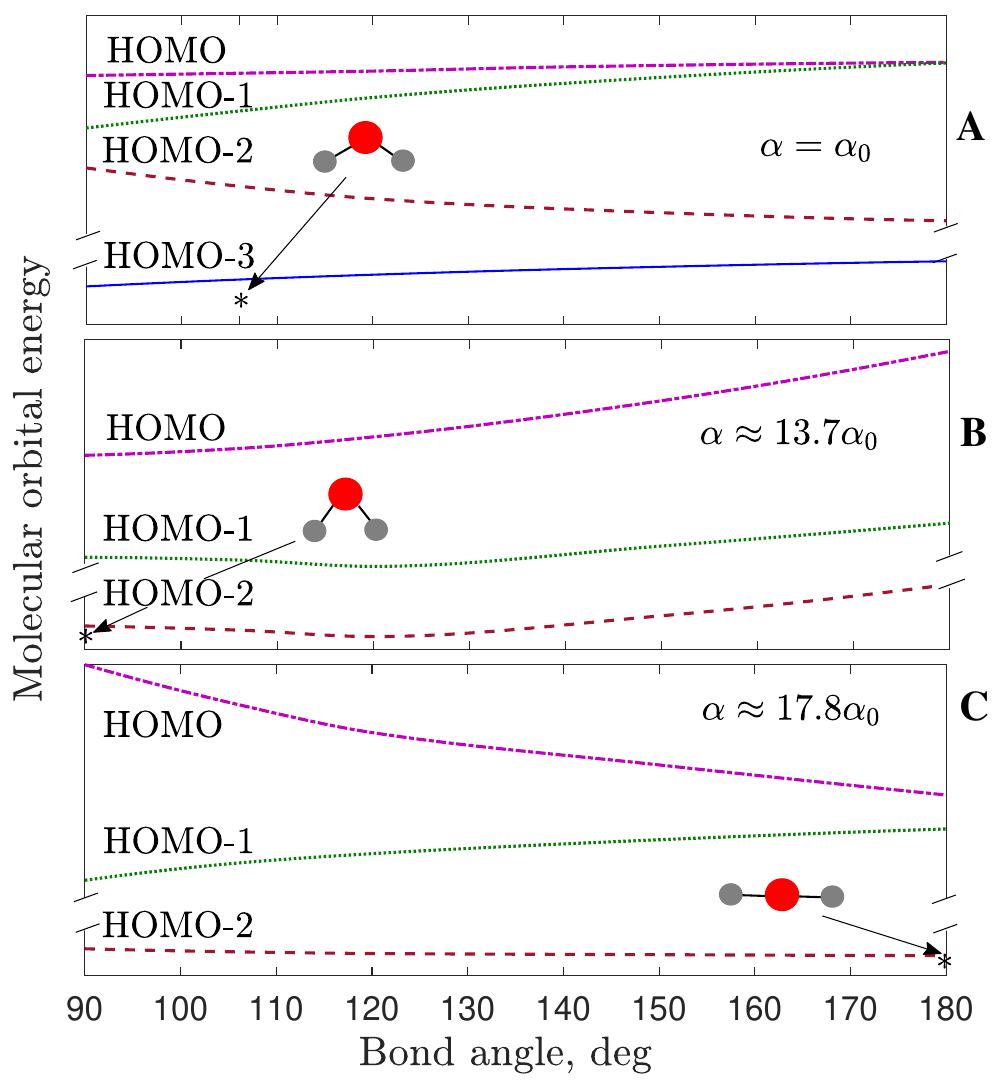}
\caption{Walsh correlation diagrams of water showing the energies of four valence molecular orbitals as functions of the bond angle at different $\alpha$. In panels B and C, the lowest occupied molecular orbital is omitted as it reduces to the $2s_{1/2}$ atomic orbital of oxygen and does not participate in the formation of oxygen-hydrogen bonds. The arrows indicate the bond angle at which the molecular geometry stabilizes. The energies of molecular orbitals at the equilibrium bond angles correspond to those in Fig.~2 A-C in the main text.}
\label{Fig:Water_Walsh}
\end{figure}

The Walsh correlation diagrams of the MO theory provide more insights into the relation between the electronic structure and geometry of the water molecule, see Fig.~\ref{Fig:Water_Walsh}. These diagrams show the energies of valence MOs as functions of the bond angle. Because the total energy of a molecule can be approximated as the sum of MO energies, the Walsh diagrams can be used to predict the values of the bond angle that minimize the total energy. At nominal $\alpha$, the interplay between the HOMO-1 and HOMO-2 energies minimizes the total energy at the bond angle of 104.5$^\mathrm{o}$. As $\alpha$ increases, the $2s_{1/2}$ orbital stabilizes and its contribution to HOMO is diminished, leading to the minima of the HOMO and total energies at the 90$^\mathrm{o}$ bond angle. At larger $\alpha$, the relativistically stabilized $3s_{1/2}$ orbital starts to contribute to HOMO, leading to the linear geometry.

\section{Dark matter clumps with large variations of fundamental constants}

In this section, we discuss in more details the properties of a dark matter (DM) clump whose interior facilitates a large variation of the fundamental constants, in particular the fine-structure constant $\alpha$. 

Depending on its mass and whether it is self-interacting, DM can exist in a variety of configurations. For example, DM in the $\sim$ GeV-TeV mass range is often searched for in the form of weakly-interacting massive particles (WIMPs) in particle detectors. In contrast, ultralight DM behaves more like classical fields due to macroscopic occupation numbers. Further, if the ultralight DM field is non-self-interacting, as in the cases of the pseudo-scalar axions or scalar dilatons/moduli, it forms a uniform background DM field oscillating coherently at the Compton frequency (``wavy'' DM). In this paper, we are concerned with ultralight DM fields which possess self-interactions and, as such, can form localized structures (``clumpy'' DM). Examples of ``clumpy" DM models include Q-balls~\cite{Coleman1985,Lee1987,Kusenko2001,KimballQball2017},
Bose stars~\cite{Hogan1988,Barranco2013,Krippendorf2018}, dark blobs~\cite{Grabowska2018}, topological defects \cite{Kibble1980,VILENKIN1985263,Vilenkin1994},
and axion quark nuggets~\cite{Zhitnitsky2019,Budker2019,sym14030459}.

The formation of DM clumps requires nonlinear self-interactions of the DM field. Nonlinearities lead to instabilities in the DM fluid, causing fluctuations of the energy density field to grow instead, thus forming clumps. Formation and properties of DM clumps have been modelled in Ref.~\cite{Brax2020}. After their formation, the clumps aggregate and afterwards play the role of the standard cold DM. In this model, the gravitationally interacting clumps behave as the prerequisite pressureless cosmological fluid. The DM field mass can span a wide range from $10^{-17}$ eV to 10 GeV. The clump's size spans a wide range, from the size of atoms ($\sim$~angstroms) to that of galactic molecular clouds ($\sim$~parsec) and its mass can vary from a milligram to thousands of solar masses. The clump mass-radius relation follows a power law, $M\sim R^n$, where $n = 3, 4, 5$, depending on the formation mechanisms and the self-interaction
potential. Because of finite-size effects, DM clumps may not be constrained by microlensing observations~\cite{niikura2019microlensing}.

For our purpose, the specific details of the DM model are unimportant and we consider a generic ultralight DM field $\phi$ that forms a pressureless ``gas'' of clumps. We only distinguish between topological defects DM (monopoles, strings, or domain walls) and nontopological solitons DM (Q-balls, oscillons, or axitons). The mass-energy of a DM clump is different in each case, leading to slightly different spatial distributions of the clumps, which determine the likelihood of such clumps encountering Earth. 

In the case of a topological defect, the DM field may be assumed to be constant, $\phi=\phi_{\rm max}$ inside the clump and zero outside. The energy density inside the clump scales as~\cite{DerPos14}
\begin{equation}
\rho\sim \phi_{\rm max}^2/d^2\,,
\end{equation}
where $d$ is the clump's transverse size. Typically, $d$ is of the order of the DM field's Compton wavelength, $d\sim\hbar/(m_\phi c)$ where $m_\phi$ is the mass of DM field. An Earth-sized clump would therefore corresponds to $m_\phi\sim10^{-14}\,{\rm eV}$. The total energy contained in a clump is thus given by
\begin{equation}\label{eq:E_TD}
E\sim \rho d^{3}=\phi_{\rm max}^2d\,,
\end{equation}
and, assuming that the clumps saturate the local DM density $\rho_{\rm DM}\approx0.3\,{\rm GeV}/{\rm cm}^3$, their number density may be written as
\begin{equation}\label{eq:number_density}
    n=\rho_{\rm DM}/E\sim\rho_{\rm DM}/(\phi_{\rm max}^2d)\,.
\end{equation}
The average time between successive ``close encounters''  with the defects is thus 
\begin{equation}\label{T_TD}
    \mathcal{T}_{\rm defect}=\frac{1}{v_gnd^2}=\frac{1}{v_g}\frac{\phi_{\rm max}^2}{\rho_{\rm DM}d}\,, 
\end{equation}
where $v_g\sim10^{-3}c\approx300$ km/s is the velocity of the clumps in the Solar system ($\sim$ Solar system velocity relative to the DM halo).

\APD{In the case of a Q-ball, the DM field inside a clump oscillates with an angular frequency $\omega$, $\phi({\bf r})=\tilde{\phi({\bf r})}e^{i\omega t}$. Also associated with each clump is a conserved charge $Q$ originated from a global $U(1)$ symmetry of the DM field~\cite{Coleman1985}
\begin{equation}\label{eq:Q_def}
 Q=\omega\int\tilde{\phi}^2d^3x\,.
\end{equation}
We consider the ``thin-wall'' approximation, $\tilde{\phi}=\phi_{\rm max}\theta(R-r)$ where $R$ is the clump radius and $\theta$ is the Heaviside function. In this approximation, Eq.~\eqref{eq:Q_def} reduces to
\begin{equation}
    Q=\omega\phi^2_{\rm max}V\,,
\end{equation}
where $V=4\pi R^3/3$ is the clump's volume. The total clump energy may be written as~\cite{Kusenko2001}
\begin{equation}\label{eq:E_tot_NS}
E=\frac{Q^2}{2V\phi^2_{\rm max}}+VU(\phi_{\rm max})\,,
\end{equation}
where $U(\phi)$ is the DM self-interaction potential. The first term in Eq.~\eqref{eq:E_tot_NS} corresponds to the clump's kinetic energy, whereas the second term corresponds to its potential energy. The volume $V$ is found by minimizing $E$. At this minimum point, one has
\begin{subequations}\label{eq:NS_energy}
\begin{align}
    V&=\frac{Q}{\phi_{\rm max}\sqrt{2U_{\rm max}}}\,,\\
    E&=\frac{Q^2}{\phi_{\rm max}U_{\rm max}}=\frac{10\pi\omega^2\phi_{\rm max}^2R^3}{3}\,.
    \end{align}
\end{subequations}

Note that in Eq.~\eqref{eq:E_tot_NS}, we have neglected, for simplicity, a surface term corresponding to the energy needed to transition from inside the Q-ball to the vacuum outside. If this term is restored, the minimization of $E$ forces the Q-ball to assume a shape with smallest surface area for a given volume, i.e., a sphere. This justifies defining the radius $R$ and volume $V$ as $V=4\pi R^3/3$.

Carrying the result~\eqref{eq:NS_energy} through as in Eqs.~\eqref{eq:number_density} and~\eqref{T_TD}, one finds the average time between successive ``close encounters''  of Earth with the Q-balls as
\begin{equation}\label{eq:T_soliton}
    \mathcal{T}_{\rm Q-ball}=\frac{5\pi}{3v_g}\frac{\omega^2d\phi^2_{\rm max}}{\rho_{\rm DM}}\,.
\end{equation}
Note that since the DM particles cannot be confined to a region smaller than their Compton wavelength $\lambda_\phi=1/m_\phi$, it is required that $d\gtrsim 1/m_\phi$, just as with topological defects. For Q-balls, their stability entails an extra requirement that $\omega\lesssim m_\phi$.}



Due to the coupling of DM to baryonic matter, an encounter with a DM clump may cause transient variations of fundamental constants~\cite{DerPos14}. For Earth-sized DM clumps with $\mathcal{T}$ of the order of a few years, these transient signals may be detected by a network of sensors searching for perturbation patterns as the clumps sweep through the network. Such direct searches are being carried out with networks of atomic clocks, such as atomic clocks aboard satellites of the Global Positioning System~\cite{Roberts2017-GPS-DM} or the trans-European network of laboratory clocks~\cite{Roberts2019-DM.EuropeanClockNetwork}.  However, if $\mathcal{T}$ reaches the cosmological scale of a few billion years, constraints on the DM parameters can  be derived via anthropic arguments presented in the main text. In some sense, the lifeforms serve as DM detectors. 

In particular, invoking the anthropic argument that a DM clump with a fine-structure constant $\alpha$ much larger than the nominal value $\alpha_0\approx1/137$ has not encountered Earth since the beginning of life on Earth is equivalent to requiring $\mathcal{T}\gtrsim 4$ Gy. Eqs.~\eqref{T_TD} and~\eqref{eq:T_soliton} then translates into
\begin{subequations}\label{phi_max}
\begin{align}
\phi_{\rm max} &\gtrsim 5\times10^6\,{\rm TeV} \quad {\rm for\,\,topological\,\,defects}\,, \label{phi_max_TD}\\
\phi_{\rm max} &\gtrsim 3\times10^7\,{\rm TeV} \quad {\rm for\,\,Q-balls}\,,\label{phi_max_NS}
\end{align}
\end{subequations}
where the bound~\eqref{phi_max_NS} was obtained by assuming that $\omega\sim m_\phi\sim1/d$.

The assumed coupling of photons to DM has the form (see main text)
\begin{equation}\label{int_DM}
    \mathcal{L}_{\rm DM-SM}\supset\phi^2F^2_{\mu\nu}/(4\Lambda^2_\gamma)\,,
\end{equation}
where $F_{\mu\nu}$ is the Faraday tensor and $\Lambda_\gamma$ is the characteristic energy scale. This coupling re-scales $\alpha$ in the core of the clump,
\begin{equation}
    \alpha_\mathrm{max}=\alpha_0\left(1-\phi^2_{\rm max}/\Lambda^2_\gamma\right)^{-1}\,.\label{alpha_alpha0}
\end{equation}
For $\alpha_\mathrm{max} \gg\alpha_0$, the regime of our interest, Eq.~\eqref{alpha_alpha0} implies that $\Lambda_\gamma\approx\phi_{\rm max}$ which, when combined with Eqs.~\eqref{phi_max}, gives the limit
\begin{subequations}\label{lambda}
\begin{align}
\Lambda_\gamma &\gtrsim 5\times10^6\,{\rm TeV} \quad {\rm for\,\,topological\,\,defects}\,, \label{lambda_TD}\\
\Lambda_\gamma &\gtrsim 3\times10^7\,{\rm TeV} \quad {\rm for\,\,Q-balls}\,,\label{lambda_NS}
\end{align}
\end{subequations}

\APD{We note that for $\Lambda_\gamma$ satisfying Eq.~\eqref{lambda}, the DM clumps are not subject to the screening by Earth's atmosphere~\cite{Stadnik2020} implied by the interaction~\eqref{int_DM}. The limit Eq.(\ref{lambda}) is largely insensitive to the specific value of $\alpha_\mathrm{max}$, resulting to the insensitivity of Eq.~\eqref{lambda} to screening effects, e.g., by the ocean water. Furthermore, Ref.~\cite{Stadnik2020} concerned the ratios $d'/d$ and $d'/R$ of the depth $d'$ at which a DM clump penetrates a region of baryonic matter to the size $d$ of the clump and the extend $R$ of the region, with ``strong screening'' happening if $d'/d\ll 1$ or $d'/R\ll 1$. For our anthropic consideration, however, the absolute value of $d'$ is more relevant. Assuming the interaction~\eqref{int_DM}, the value of the screening depth $d'$ is given by~\cite{Stadnik2020}
\begin{equation}
    d'=\Lambda_\gamma/\sqrt{2\rho_\gamma}\,,
\end{equation}
where $\rho_\gamma$ is the electromagnetic mass-energy density of the screening matter. Using $\Lambda_\gamma\sim10^7$ TeV, as well as $\rho^{\rm water}_\gamma\approx9.2\times10^{-4}\,{\rm g/cm}^3$ and $\rho^{\rm rock}_\gamma\approx8.8\times10^{-3}\,{\rm g/cm}^3$~\cite{Stadnik2020}, one finds that $d'_{\rm water}\sim10\,{\rm km}$ and $d'_{\rm rock}\sim3\,{\rm km}$. The value $d'_{\rm water}\sim10\,{\rm km}$ is similar to the depth of the Mariana Trench, while the value $d'_{\rm rock}\sim3\,{\rm km}$, although small compared to the size of Earth, is also large in comparison with the soil depths at which life-forms may be found.}

Let us now consider the implications of Eqs.~\eqref{phi_max}. In the following discussion, we shall assume that $\omega\sim m_\phi\sim1/d$. Under these assumptions, Eqs.~\eqref{eq:E_TD} and~\eqref{eq:NS_energy} are the same up to a numerical factor. As a result, the results below apply to both topological- defect and nontopological-soliton DM. 

Assuming an Earth-sized clump, $d\approx\,13,000\,{\rm km}$, the energy density inside the clump is found to be
\begin{equation}\label{eq:rho_clump}
    \rho\sim 10^{-6}\,{\rm kg}/{\rm cm}^3\,,
\end{equation}
and its total mass-energy is
\begin{equation}
\begin{aligned}
     E&\sim 10^{-9}\text{ mass of Earth}\,.
\end{aligned}
\end{equation}
We observe that an Earth-sized DM clump with $\phi_{\rm max} \sim 5\times10^6\,{\rm TeV}$ only weights as much as a small asteroid. In particular, such a clump does not collapse into a black hole. Further, if such a clump enters the solar system, its presence does not cause significant disturbances to the planetary orbits therein. Additionally, galactic structure formation would occur as per conventional cold dark matter theory~\cite{blumenthal1984,Brax2020}.

On the other hand, the number of these Earth-sized clumps could be quite large. Indeed, assuming that the DM clumps saturate the local DM density, we may estimate their number density $n$ as
\begin{equation}
    n\sim  {10}^{11}/\text{light-year}^3\,,
\end{equation}
which is to be compared with the stellar number density in the Milky Way $4\times10^{-3}/\text{light-year}^3$. Correspondingly, the total number of clumps in the Milky Way is
\begin{align}
    N&=nV_{\rm Milky Way}\approx 8\times {{10}^{12}}\,\left( \text{light-year} \right)^3\nonumber\\
    &\times {{10}^{11}}/\text{light-year}^3\approx {{10}^{24}}\,,
\end{align}
much larger than the total number of stars in the Milky Way, $\sim 4\times 10^{11}$.

The average distance $L$ between two clumps may be estimated as
\begin{equation}\label{eq:L}
    L\sim1/\sqrt[3]{n}\approx 10^9\,\text{ km}\,,
\end{equation}
which is the same order as the size of the Solar system.



\APD{\section{Effect of $\alpha$-variation on proton stability}
As mentioned in the main text, increasing the value of $\alpha$ leads to proton decay. In this appendix, we consider the decay of a proton into a neutron by emitting a positron
\begin{equation}\label{eq:beta+decay}
    \beta^+\,\text{decay}:\quad p\rightarrow n + e^+ +\nu_e\,,
\end{equation}
or by capturing an atomic electron
\begin{equation}\label{eq:eq_ecap}
    \text{orbital electron capture:}\quad p+e^-\rightarrow n +\nu_e\,.
\end{equation}
Both processes are accompanied by the emission of an electron neutrino. 

In a $\beta^+$-decay process, Eq.~\eqref{eq:beta+decay}, the energy conservation condition (assuming that the proton is initially at rest) reads
\begin{equation}\label{eq:beta+cond}
    m_p=m_n+E_R+E_{e^+}+E_\nu\,,
\end{equation}
where $m_p$ and $m_n$ are the proton and neutron masses, $E_R$ is the neutron recoil energy, $E_{e^+}$ is the outgoing positron energy, and $E_{\nu}$ is the neutrino energy. In the current consideration, it is sufficient to assume that the neutrino is massless, so $E_\nu\approx p_\nu$, where $p_\nu$ is the neutrino momentum. 
Similarly, the energy conservation condition for orbital electron capture, Eq.~\eqref{eq:eq_ecap}, reads
\begin{equation}\label{eq:elecccond}
    m_p+m_e-\varepsilon_e=m_n+E_R+E_{\nu}\,,
\end{equation}
where $\varepsilon_e>0$ is the electron removal energy. For the $1s_{1/2}$ electron, $\varepsilon_e=-\varepsilon_{1s_{1/2}}$ introduced in the main text. Neglecting the neutrino mass, the conditions~\eqref{eq:beta+cond} and~\eqref{eq:elecccond} imply
\begin{subequations}\label{eq:tot_conds}
    \begin{align}
        m_p-m_n&\geq m_e\quad\text{for }\beta^+\text{ decay}\,,\label{cond:betadecay2}\\
        m_p-m_n&\geq \varepsilon_e-m_e\quad\text{for orbital electron capture}\label{cond:ecapture2}\,,
    \end{align}
\end{subequations}
which show that at nominal $\alpha$, both $\beta^+$ decay and orbital electron capture are forbidden for free proton and the hydrogen nucleus.

The conditions~\eqref{eq:tot_conds} may be satisfied, however, as $\alpha$ increases. The mass difference $\Delta m\equiv m_n-m_p$ is due the down-up quark mass difference $m_d-m_u$, the gluons interaction energy, and the electromagnetic interaction energy between quarks. To leading order, one expects the difference $m_d-m_u$ and the gluons interaction to be independent of $\alpha$ while the quark-quark electromagnetic energy is $\propto\alpha$. As a result, we write, to leading order
\begin{equation}\label{eq:del m alpha}
    \Delta m = A+B\alpha\,,
\end{equation}
where $A$ and $B$ are constants. Indeed, Eq.~\eqref{eq:del m alpha} is supported by lattice QCD calculations~\cite{Borsanyi2015,Romiti:2022DY}. In particular, Ref.~\cite{Borsanyi2015} found that for $\alpha\approx2\alpha_0$, $\Delta m\approx m_e$, thus opening up the electron capture decay channel for hydrogen atom (see Eq.~\ref{cond:ecapture2}). This fact allows one to determine the constants $A$ and $B$, arriving at
\begin{equation}\label{eq:deltam=fa}
    \Delta m =\Delta m_0-\frac{\Delta m_0-m_e}{\alpha_0}(\alpha-\alpha_0)\,,
\end{equation}
where $\Delta m_0\approx1.29$ MeV is the neutron-proton mass difference at nominal $\alpha$. 
It may be observed from Eq.~\eqref{eq:deltam=fa} that for $\alpha/\alpha_0\gtrsim 3.3$, the condition~\eqref{cond:betadecay2} is satisfied and $\beta^+$ decay happens in hydrogen. Let us now estimate the rates of $\beta^+$ decay and electron capture in hydrogen at large $\alpha$. 

The $\beta^+$ decay rate is given by~\cite{wu1966,heyde2020basic}
\begin{equation}\label{eq:rate_beta+}
    w_{\beta^+}=\frac{g^2m^5_e}{2\pi^3}f\left(E^{\rm max}_e/m_e\right)\left|M_{p\rightarrow n}\right|^2\,,
\end{equation}
where $g=8.8\times10^{-5}\text{ MeV fm}^3$ is the $\beta$-decay strength constant, $M_{p\rightarrow n}$ is the nuclear matrix element, the function $f(x)$ is given by
\begin{equation}\label{eq:def_f}
    f(x)=\int_1^x\sqrt{w^2-1}\left(x-w\right)^2wdw\,,
\end{equation}
and $E^{\rm max}_e$ is the maximum electron energy determined by setting $E_R=E_\nu=0$ in Eq.~\eqref{eq:beta+cond}, i.e., $E^{\rm max}_e=-\Delta m$. 

For electron capture by the proton in hydrogen, the rate is given by~\cite{wu1966,heyde2020basic}
\begin{equation}\label{eq:rate_ec}
    w_{\rm ec}=\frac{g^2}{\pi}\rho_eE^2_{\nu}\left|M_{p\rightarrow n}\right|^2\,,
\end{equation}
where $\rho_e$ is the electron density at the proton. The neutrino energy $E_\nu$ may be related to $\Delta m$ by using Eq.~\eqref{eq:elecccond} and the conservation of momentum $p_R=-p_\nu\approx-E_\nu$, where $p_R$ is the neutron recoil momentum
\begin{equation}\label{eq:Tv_eq}
    \frac{E_\nu^2}{2m_n}+E_\nu=-\Delta m+m_e-\varepsilon_e\,.
\end{equation}
Evidently, Eq.~\eqref{eq:Tv_eq} has a positive solution $E_\nu$ only if its right hand side is positive, i.e., that the condition~\eqref{cond:ecapture2} is satisfied. Note that we have used the nonrelativistic form of the nuclear recoil energy $E_R=p_R^2/(2m_n)\approx E_\nu^2/(2m_n)$. Numerical calculations show that for $\alpha \lesssim 20\alpha_0$, $E_\nu\lesssim 14\text{ MeV}\ll m_n$, thus justifying the use of the nonrelativistic formula. 

The electron density $\rho_e(r)$ may be estimated using the Dirac wave function, anticipating the importance of relativistic effects on the bound electron as $\alpha$ is increased. For a $1s_{1/2}$ electron, one has
\begin{equation}
    \rho_e(r)=\frac{(2\alpha m_er)^{2\gamma+1}}{2(\gamma+1)\Gamma(2\gamma+1)}\left[(\gamma+1)^2+\alpha^2\right]\frac{e^{-2\alpha m_er}}{r^3}\,,
\end{equation}
where $\gamma=\sqrt{1-\alpha^2}$ and $\Gamma(x)$ is the gamma function. For a point-like nucleus, the density $\rho_e(r)$ tends to infinity as $r\rightarrow0$. This divergence may be avoided by considering the finite size of the nucleus. For our estimate, we simply choose a cut-off at the nuclear radius $R_N$ and use $\rho_e(R_N)$ in Eq.~\eqref{eq:rate_ec}. For hydrogen, $R_N\approx1$ fm. 

To estimate the rates $w_{\beta^+}$ and $w_{\rm ec}$, we also need the squared nuclear matrix element $\left|M_{p\rightarrow n}\right|^2$, which is conventionally separated in to a Fermi (F) part and a Gamow-Teller (GT) part~\cite{wu1966,heyde2020basic}
\begin{subequations}\label{eq:FermiGamowTeller}
\begin{align}
    \left|M_{p\rightarrow n}\right|^2&=\left|M_{np}^F\right|^2+\lambda^2\left|M_{np}^{GT}\right|^2\,,\\
    \left|M_{np}^F\right|^2&\equiv\left|\int\psi^\dagger_nQ\psi_pd{\bf r}\right|^2\,,\\
    \left|M_{np}^{GT}\right|^2&\equiv\sum_{\mu=-1}^1\left|\int\psi^\dagger_nQ\sigma_\mu\psi_pd{\bf r}\right|^2\,,
\end{align}
\end{subequations}
where $Q$ is the operator changing $p$ to $n$, $\sigma_\mu$ is the spherical basis component of the Pauli matrix, and $\lambda=1.259$. The form of the Fermi matrix element implies that it involves no nucleon spin flip, $\Delta I=0$, wheres a Gamow-Teller transition involves a nuclear spin change $\Delta I=0,\pm 1$. In particular, for a pure Fermi transition, $\left|M_{p\rightarrow n}\right|^2=\left|M_{np}^F\right|^2=1$, whereas for a pure Gamow-Teller transition with $\Delta I=1$, $\left|M_{p\rightarrow n}\right|^2=\lambda^2\left|M_{np}^{GT}\right|^2=3\lambda^2$. Since we are interested in the transition rate, we sum over the final neutron spin states, so that $\left|M_{p\rightarrow n}\right|^2=1+3\lambda^2\approx5.8$. 

It is also useful to consider the ratio $w_{\beta^+}/w_{\rm ec}$, which depends only on the kinematics of the two decay processes. Indeed, from Eqs.~\eqref{eq:rate_beta+} and~\eqref{eq:rate_ec}, one finds
\begin{equation}
    \frac{w_{\beta^+}}{w_{\rm ec}}=\frac{f\left(-\Delta m/m_e\right)}{2\pi^2}\frac{m_e^5}{\rho_eE_\nu^2}\approx\frac{f\left(\xi\right)}{2\pi^2\xi^2}\frac{m_e^3}{\rho_e}\,,
\end{equation}
where we have used the fact that $E_\nu\approx-\Delta m$ for large $\alpha$, see Eq.~\eqref{eq:Tv_eq}. We have also defined $\xi\equiv-\Delta m/m_e$. The ratio $f(\xi)/\xi^2$ grows very fast with respect to $\xi$, which, from Eq.~\eqref{eq:deltam=fa}, increases linearly with $\alpha$. With $\alpha\approx 20\alpha_0$, $f(\xi)/\xi^2\approx4\times10^5$. The ratio $m_e^3/\rho_e$ decays quickly with increasing $\alpha$ but its value remains large, $\sim 10^4-10^2$, for $\alpha\lesssim20\alpha_0$. As a result, for large $\alpha$, $\beta^+$ decay is the dominant channel for the $p\rightarrow n$ transmutation.

In Fig.~\ref{fig:rates}, we plot the proton-to-neutron $\beta^+$ decay and electron capture rates for $0\leq\alpha/\alpha_0\leq20$. It is indeed clear from the plot that for large $\alpha$, $\beta^+$ decay dominates. Furthermore, we observe that for $\alpha\approx5.5\alpha_0$, $w_{\beta^+}\approx0.03\,{\rm s}^{-1}$, corresponding to lifetime of $\sim30$ s. We note also that at $\alpha\approx5.5\alpha_0$, the neutron recoil energy is $E_R\approx 4$ keV, which, although much smaller than the neutron rest mass and thus allowing treating the recoil neutron nonrelativistically, is large enough to cause irreversible scattering and ionization in the surrounding media.}
\begin{figure}[h!]
    \centering
    \includegraphics[width=3in]{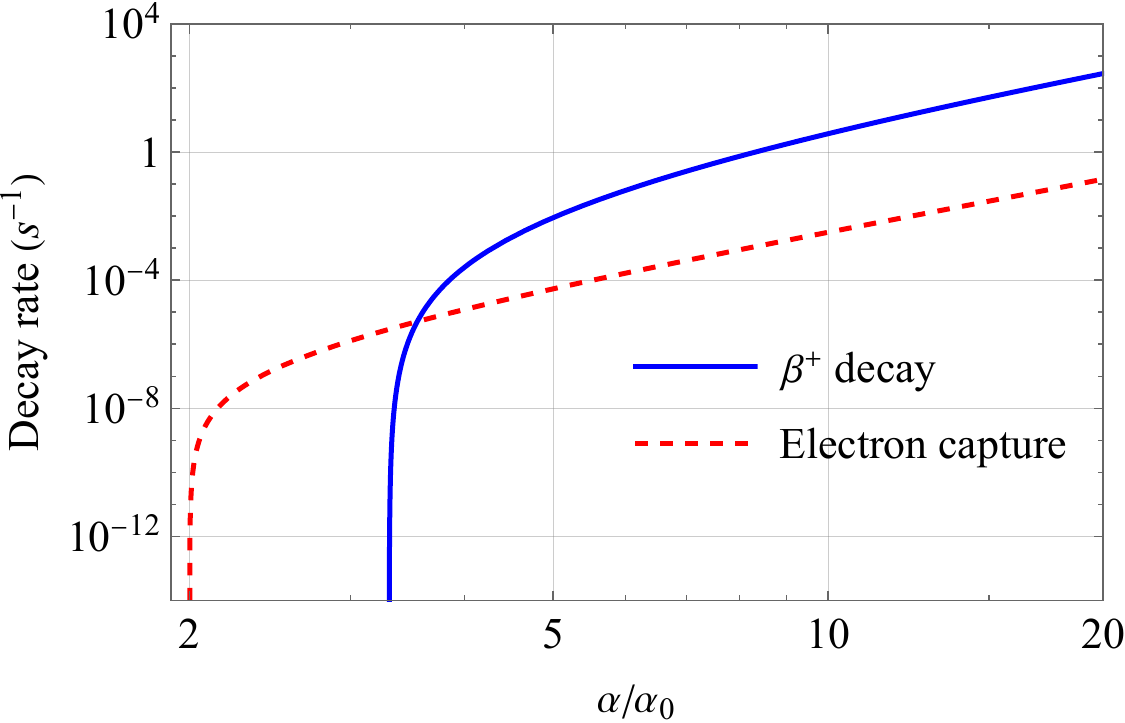}
    \caption{Dependence of the proton-to-neutron $\beta^+$ decay rate and the electron capture rate on $\alpha$.}
    \label{fig:rates}
\end{figure}

\bibliographystyle{apsrev4-2}
\bibliography{bao}
\end{document}